\documentclass[useAMS]{emulateapj}

\usepackage{latexsym,graphicx,natbib,apjfonts}
\usepackage{amsmath}

\newcommand\simless{{\thinspace \rlap{\raise 0.5ex\hbox{$\scriptstyle  {<}$}}
    {\lower 0.3ex\hbox{$\scriptstyle  {\sim}$}} \thinspace }}  
\newcommand\simgreat{{\thinspace \rlap{\raise 0.5ex\hbox{$\scriptstyle  {>}$}}
    {\lower 0.3ex\hbox{$\scriptstyle  {\sim}$}} \thinspace }}  
\newcommand\msun{\, \rm M_\odot} 
\newcommand\kms{{\, \rm km\,s^{-1}}}
\newcommand\mbh{M_{\rm BH}}
\newcommand\vbh{V}
\newcommand\mg{M_{\rm gal}}
\newcommand\rb{r_{\rm b}}
\newcommand\re{R_{\rm e}}
\newcommand\rhob{\rho_{\rm b}}
\newcommand\vk{V_{\rm kick}}
\newcommand\ve{V_{\rm esc}}
\newcommand\beq{\begin{equation}}
\newcommand\eeq{\end{equation}}

\newcommand\tdf{T_{\rm df}}
\newcommand\form{F}
\newcommand\mdef{M_{\rm def}}
\newcommand\mgal{M_{\rm gal}}

\shortauthors{Gualandris and Merritt}
\shorttitle{Ejection of Supermassive Black Holes from Galaxy Cores}

\begin{document}

\title{Ejection of Supermassive Black Holes from Galaxy Cores}

\author{Alessia Gualandris and David Merritt}

\affil{Center for Computational Relativity and Gravitation, 
  Rochester Institute of Technology, 
  78 Lomb Memorial Drive, Rochester, NY 14623}

\email{alessiag,merritt@astro.rit.edu}

\begin{abstract}
  Recent numerical relativity simulations have shown that the emission
  of gravitational waves during the merger of two supermassive black
  holes (SMBHs) delivers a kick to the final hole, with a magnitude as
  large as $4000\kms$.  We study the motion of SMBHs ejected from
  galaxy cores by such kicks and the effects on the stellar
  distribution using high-accuracy direct $N$-body simulations.
  Following the kick, the motion of the SMBH exhibits three distinct
  phases.  (1) The SMBH oscillates with decreasing amplitude, losing
  energy via dynamical friction each time it passes through the core.
  Chandrasekhar's theory accurately reproduces the motion of the SMBH
  in this regime if $2\lesssim\ln\Lambda\lesssim 3$ and if the
  changing core density is taken into account.  (2) When the amplitude
  of the motion has fallen to roughly the core radius, the SMBH and
  core begin to exhibit oscillations about their common center of
  mass. These oscillations decay with a time constant that is at least
  $10$ times longer than would be predicted by naive application of
  the dynamical friction formula.  During this phase, the SMBH is
  typically displaced from the peak of stellar density by roughly the
  core radius.  (3) Eventually, the SMBH reaches thermal equilibrium
  with the stars.  We use straightforward scaling arguments to
  estimate the time for the SMBH's oscillations to damp to the
  Brownian level in real galaxies and infer times as long as $\sim
  1\,\rm Gyr$ in the brightest galaxies.  The longevity of the
  oscillations makes this mechanism competitive with others that have
  been proposed to explain double or offset nuclei.  Ejection of SMBHs
  also results in a lowered density of stars near the galaxy center;
  mass deficits as large as five times the SMBH mass are produced for
  kick velocities near the escape velocity.  We compare the $N$-body
  density profiles with luminosity profiles of early-type galaxies in
  Virgo and show that even the largest observed cores can be
  reproduced by the kicks, without the need to postulate
  ``hypermassive'' binary SMBHs.  Implications for displaced AGNs and
  helical radio structures are discussed.
\end{abstract}


\section{Introduction}
\label{sec:intro}

The recent breakthroughs in numerical relativity
\citep{Pretorius:05,Campanelli:06,Baker:06a} have allowed a number of
groups to evolve binary black holes (BHs) to full coalescence.  The
final inspiral is driven by emission of gravitational waves, and in
typical (asymmetric) inspirals, a net impulse is imparted to the
system due to anisotropic emission of the waves
\citep{Bekenstein:73,Fitchett:84,Favata:04}.  Early arguments that the
magnitude of the recoil velocity would be modest for non-spinning BHs
\citep{Redmount:89} were confirmed by the simulations, which found
$\vk\lesssim 200\kms$ in the absence of spins
\citep{Baker:06b,Gonzalez:07a,Herrmann:07}.  The situation changed
dramatically following the first simulations of ``generic'' binaries,
i.e., binaries in which the individual BHs were spinning and in which
the spins were allowed to have arbitrary orientations
\citep{Campanelli:07a}.  Kicks as large as $\sim 2000\kms$ have now
been confirmed \citep{Campanelli:07b,Gonzalez:07b,Tichy:07}, and
simple scaling arguments suggest that the maximum kick velocity would
probably increase to $\sim 4000\kms$ in the case of maximally-spinning
holes \citep{Campanelli:07b}.  The most propitious configuration for
the kicks consists of an equal-mass binary in which the individual
spin vectors are oppositely aligned and oriented parallel to the
orbital plane.  The kick amplitude also depends sensitively on the
angle between the BH spin vectors and their linear momenta shortly
before the plunge \citep{Lousto:07}.

Galaxy escape velocities are $\lesssim 3000\kms$ \citep{MMFHH:04},
which means that gravitational wave recoil can in principle displace
coalescing supermassive black holes (SMBHs) arbitrarily far from
galaxy centers, or even eject them completely.  The actual
distribution of kick velocities is very uncertain, since it depends on
the unknown distribution of binary mass ratios and spins, but most
kicks are probably $\lesssim 10^3\kms$.  A SMBH that is kicked with
less than escape velocity will travel some maximum distance from the
galaxy center after which its orbit decays due to dynamical friction;
most of the energy loss takes place during passages through the galaxy
center.  Removal of the SMBH from the core has the effect of
transferring kinetic energy to the stars and lowering the core density
\citep{Redmount:89,MMFHH:04,Boylan:04}.  This implies a more gradual
return of the SMBH to a zero-velocity state than in a galaxy with
fixed density.

In fact, however, the SMBH is not expected to ever reach a state
of zero kinetic energy.
When its energy falls to a value

\begin{equation}
\frac{1}{2} \mbh V^2 \approx \frac{1}{2} m_{\star} {\rm v}_{\star}^2
\label{eq:vbrown}
\end{equation}

with respect to the galaxy central potential, where $m_\star$ and
${\rm v}_\star$ are a typical stellar mass and velocity respectively,
random gravitational perturbations from stars act to accelerate the
SMBH as often as they decelerate it.  This is the regime of
gravitational Brownian motion \citep{Young:77,BW:76,MBL:07}.  A
natural definition of the ``return time'' of a kicked SMBH is the time
required for dynamical friction to reduce the SMBH's mean kinetic
energy to the Brownian value.  Applying standard expressions for the
dynamical friction force leads one to the conclusion that this would
occur in a relatively short time, of order a few orbital periods,
after dynamical friction has returned the kicked SMBH to the core.

The $N$-body simulations presented here were designed to test these
expectations by evaluating the return times of kicked SMBHs and by
quantifying the induced changes in galaxy structure.  These processes
can not be studied accurately using classical dynamical friction
theory since the SMBH substantially modifies the core as it recoils
and falls back.  Approximate $N$-body schemes, e.g. tree or grid
codes, are also not well suited to the problem since they can not
robustly follow both the early (collisionless) and late (collisional)
evolution of the SMBH.  Large particle numbers are required in
order to cleanly separate the collisional and collisionless regimes.

These various requirements can currently be met only with parallel,
direct $N$-body codes running on special-purpose supercomputers.  Our
simulations use the $\phi$GRAPE integrator \citep{Harfst:07} as
implemented on {\tt gravitySimulator}, a 32-node supercomputer
employing GRAPE-6A accelerator boards \citep{GRAPE-6A}.

Our findings are surprising in one important respect.  After returning
to the core, the kicked SMBH exhibits long-lived oscillations with
amplitude comparable to the core radius\footnote{A movie showing the
oscillations is available at
http://ccrg.rit.edu/Research/Publications.php?paper=0708.0771.}.
These oscillations eventually decay but with a time constant that is
at least an order of magnitude longer than would be predicted by a
straightforward application of the dynamical friction equation.  We
demonstrate that the existence, amplitude and damping time of these
oscillations are independent of the number $N$ of ``star'' particles
used in the simulations, for $N$ up to $2\times 10^6$.  The
oscillations are similar to those first reported by R. Miller and
collaborators \citep{Miller:92,Miller:96} in their pioneering $N$-body
studies of the central regions of galaxies.  A number of other authors
have reported low effective values of the dynamical friction force as
it acts on massive objects that inspiral into constant-density cores
\citep{Bontekoe:88,Bertin:03,Read:06} or on rotating bars
\citep{Weinberg:02,Klypin:03}.  Our use of a high-accuracy,
direct-summation $N$-body code combined with large particle numbers
greatly reduces the possibility that our results are an artifact of
the potential calculation scheme, an issue that has plagued the
interpretation of similar results in the past \citep{Zaritsky:88}.

\S\,\ref{sec:models} describes the initial models and the $N$-body
algorithm.  Evolution of the SMBH's orbit is described in detail in
\S\,\ref{sec:bhmotion}, and the induced changes in galaxy structure
are described in \S\,\ref{sec:profiles}, where the $N$-body models are
compared to luminosity profiles of core galaxies.  \S\,\ref{sec:times}
presents estimates of the SMBH return times in real galaxies, and
\S\,\ref{sec:obs} discusses some of the observable consequences of the
kicks.

\section{Initial models and numerical methods}
\label{sec:models}

The light profiles of elliptical galaxies and the bulges of spiral
galaxies are generally well described in terms of the S{\'e}rsic model
\citep{1963BAAA....6...41S, 1968adga.book.....S}, which is a
generalization of the \cite{1948AnAp...11..247D,1959HDP....53..275D}
law.  The most luminous elliptical galaxies depart systematically from
the S{\'e}rsic law near the center, where they show evidence for
partially depleted stellar cores
\citep{Faber:97,Milos:02,Graham:04,ACS:VI}.  Formation of a binary
SMBH following a galaxy ``major merger'' has been shown to produce
cores of roughly the right magnitude \citep{MM:01,Merritt:06},
although some observed cores are too large to be easily explained by
this model (a point we return to in detail below).

As approximate representations of galaxies with binary-depleted cores,
we adopt core-S{\'e}rsic models \citep{Graham:03} for our initial conditions.
The {\it space} density profile of a galaxy that follows 
the core-S{\'e}rsic law in projection
can be accurately approximated as \citep{2005MNRAS.362..197T} 
\begin{eqnarray}
\rho \left(r\right) & = & \rho^{'} \left[1 + \left(\frac{\rb}{r}\right)^{\alpha}\right]^{\gamma/\alpha}\nonumber \\
                    &   & \left[\left(r^{\alpha} + \rb^{\alpha}\right)/\re^{\alpha}\right]^{-p/\alpha} 
e^{-b \left[\left(r^{\alpha}+\rb^{\alpha}\right)/\re^{\alpha}\right]^{1/n \alpha}}
\label{eq:PS1}
\end{eqnarray}
with
\begin{equation}
\rho^{'} = \rhob \,2^{\left(p-\gamma\right)/\alpha} \left(\frac{\rb}{\re}\right)^p 
e^{b \left(2^{1/\alpha} \rb/\re \right)^{1/n}}.
\label{eq:PS2}
\end{equation}
Equation~(\ref{eq:PS1}) is a modification of the Prugniel-Simien model
\citep{PS:97}.  Here, $\re$ is the effective (half-mass) radius of the
projected galaxy; $\rb$ is the break (core) radius; $\rhob$ is the
space density at $r=r_b$; and $\alpha$ regulates the sharpness of the
transition from core to outer profile.  The parameter $n$ describes
the curvature of the S{\'e}rsic profile and $b$ and $p$ are fixed
functions of $n$ \citep{PS:97,2005MNRAS.362..197T}.  Monte-Carlo
initial conditions were generated using the scheme of \cite{Szell:05},
after including the gravitational potential of a central point
particle representing the SMBH.

The parameters used for our initial models are listed in
Table~\ref{tab:models}. The table also reports names for the
different runs based on the adopted ratio of SMBH mass to galaxy mass
and initial core radius.  Core radii were chosen so as to give initial
mass deficits of roughly $\mbh$, as observed for the majority of
luminous early-type galaxies \citep{Merritt:06}.  We note that
$\gamma=0.5$ is the shallowest power-law profile that is consistent
with a non-negative, isotropic distribution of stellar velocities
around the BH.

\begin{table}
\begin{center}
\caption{Parameters of the initial models.}
\label{tab:models}
\begin{tabular}{cccccc}
\hline 
name & $n$ & $\alpha$ & $\rb$ & $\gamma$ & $\mbh/\mg$\\
\hline 
A1 & 4.0 & 2.0 & 0.014   & 0.55 & $1.0\times 10^{-3}$ \\ 
A2 & 4.0 & 2.0 & 0.0095  & 0.55 & $1.0\times 10^{-3}$ \\ 
B & 4.0 & 2.0 & 0.027   & 0.55 & $3.0\times 10^{-3}$ \\ 
\hline
\end{tabular}
\end{center}
\end{table}

The initial models were evolved using the $\phi$GRAPE numerical
integrator \citep{Harfst:07}.  This direct-summation code employs a
fourth-order Hermite integrator with predictor-corrector scheme and
hierarchical time steps.  The MPI parallelization strategy is designed
to minimize the amount of communication among different computing
nodes and to make efficient use of the special-purpose GRAPE hardware.
All the simulations presented in this work were performed on the
32-node cluster {\tt
gravitySimulator}\footnote{http://wiki.cs.rit.edu/bin/view/GRAPEcluster}
at the Rochester Institute of Technology.  Most of our simulations
used $N=0.5\times 10^6$ equal-mass particles to represent the galaxy
although some runs used larger $N$.  We set the ratio of BH mass to
galaxy mass, $\mbh/M_{\rm gal}$, to be $(1,3)\times 10^{-3}$, typical
for observed galaxies \citep{2001MNRAS.320L..30M}.  For each model
described in Table~\ref{tab:models}, we chose eleven different values
of the kick velocity $\vk$ in units of the central escape speed $\ve$:
$\vk=(0.1,0.2,..,1.1) \times\ve$; the latter was computed numerically
from the initial $N$-body models.  In order to guarantee energy
conservation, we used a time-step accuracy parameter $\eta = 0.01$.
This ensures a relative energy error smaller than one part in $10^6$.
An accuracy parameter twice as big would approximately halve the
integration time but would result in a relative energy error of
$10^{-4}$, which we do not consider acceptable for this study.  A
softening length $\epsilon = 10^{-4}$ was assigned to both the stars
and the BH.  Such a small softening length has been shown not to
affect even the Brownian motion of a massive particle in models like
ours \citep{MBL:07}.

Throughout the paper we adopt units according to which the
gravitational constant $G$, the effective radius $\re$ in equation
(\ref{eq:PS1}), and the total galaxy mass $M_{\rm gal}$ are unity.
The models can be scaled to physical units as follows:
\begin{mathletters}
\begin{eqnarray}
\left[T\right] &=& \left(\frac{G\,M_{\rm gal}}{\re^3}\right)^{-1/2} \\
&=& 7.75 \times 10^6 {\rm yr} 
\left(\frac{M_{\rm gal}}{10^{11}\msun}\right)^{-1/2}
\left(\frac{\re}{3\,\rm kpc}\right)^{3/2},\\
\left[V\right] &=& \left(\frac{G M_{\rm gal}}{\re}\right)^{1/2} \\
&=& 378\kms
\left(\frac{M_{\rm gal}}{10^{11}\msun}\right)^{1/2}
\left(\frac{\re}{3\,\rm kpc}\right)^{-1/2}. 
\end{eqnarray}
\label{eq:units}
\end{mathletters}

\section{The black hole motion}
\label{sec:bhmotion}

\subsection {General Remarks}
\begin{figure}
  \begin{center}
      \includegraphics[width=8.0cm]{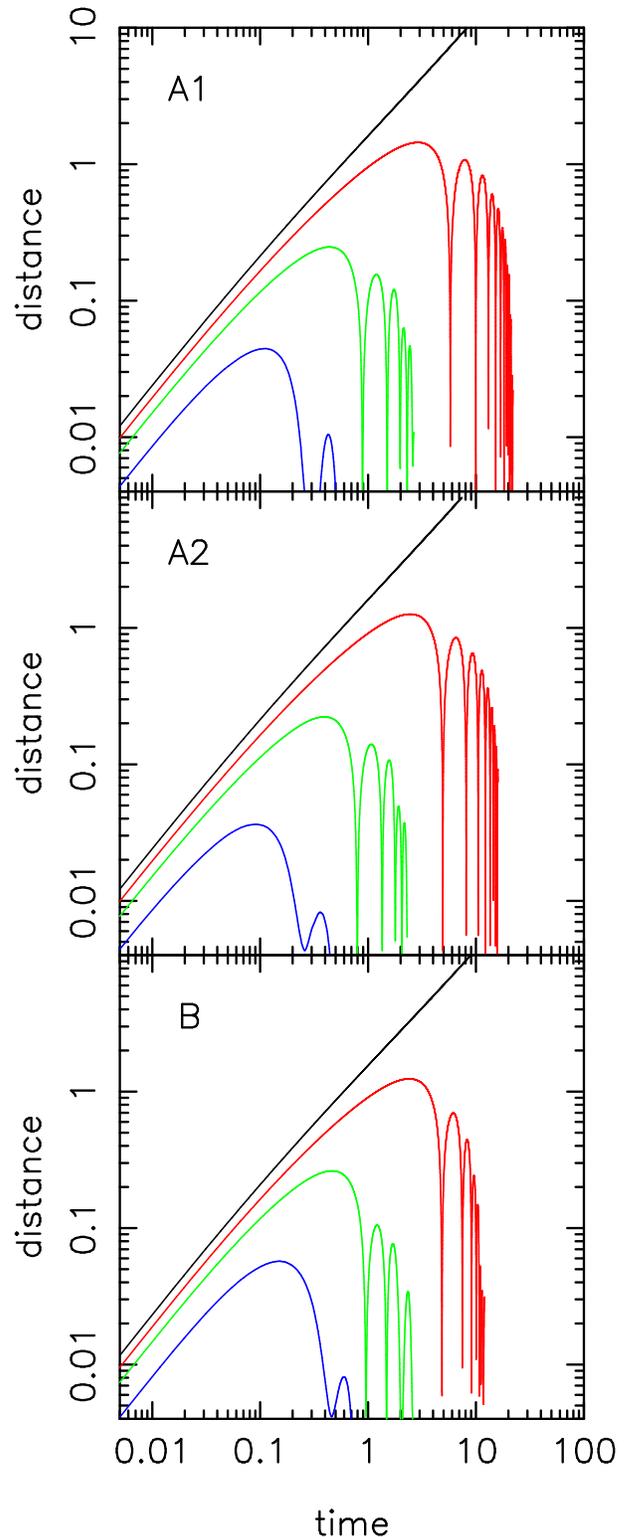}
  \end{center}
  \caption{BH trajectories in models A1, A2 and B, for $\vk/\ve=0.4$
    (blue/lower), $0.7$ (green), $0.9$ (red) and $1.1$ (black).}
  \label{fig:roft}
\end{figure}
Figure~\ref{fig:roft} (which can be compared with Fig.~1 of
\citealp{madauq:04} shows BH trajectories in models A1, A2 and B for
$\vk/\ve=(0.4,0.7,0.9,1.1)$.  For $\vk \ge \ve$ the black hole escapes
the galaxy on an unbound orbit.
\begin{figure}
  \begin{center}
    \includegraphics[width=8.5cm]{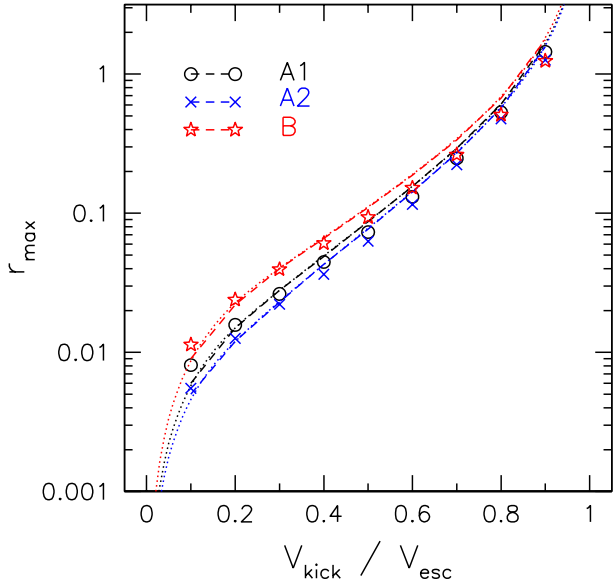}
  \end{center}
  \caption{Maximum displacement of the BH from the galaxy center.  The
    data points show the results from the simulations while the lines
    are estimates in the absence of dynamical friction.  The dashed
    lines represent numerical estimates from the computation of the
    potential of the $N$-body system at time $t = 0$ while the dotted
    lines represent theoretical estimates from the analytic expression
    of the potential in a core-S{\'e}rsic model.}
  \label{fig:rmax}
\end{figure}
The maximum displacement of the BH ($r_{\rm max}$) is shown in
Figure~\ref{fig:rmax}.  The data from the simulations (points) are
compared to theoretical (dotted lines) and numerical (dashed lines)
estimates of $r_{\rm max}$ in the absence of dynamical friction.  The
theoretical and numerical estimates are obtained from the initial
$N$-body data by assuming conservation of total energy for the BH:
$r_{\rm max}$ is the distance at which the gravitational potential of
the system equals the initial total energy of the BH. For the
theoretical solution we use the expression of the potential in a
core-S{\'e}rsic model \citep[see equations 7 through 13
of][]{2005MNRAS.362..197T} while for the numerical solution we compute
the potential at different radii from the $N$-body data.
The two estimates are for practical purposes indistinguishable.

Dynamical friction affects the maximum displacement of the BH only for
moderately large kicks, where the data points appear systematically
lower than the theoretical curves. Values of $r_{\rm max}$ larger than
the expected turning points in the first orbit are due to the 
rapidly-expanding core.

During the initial outward journey, dynamical friction does not
strongly influence the motion of the BH, and the maximum displacement
is similar to that of an energy-conserving orbit.  We note that a kick
velocity larger than about $0.3\,\ve$ is necessary to bring the BH
beyond the core.
\begin{figure}
  \begin{center}
    \includegraphics[width=8.5cm]{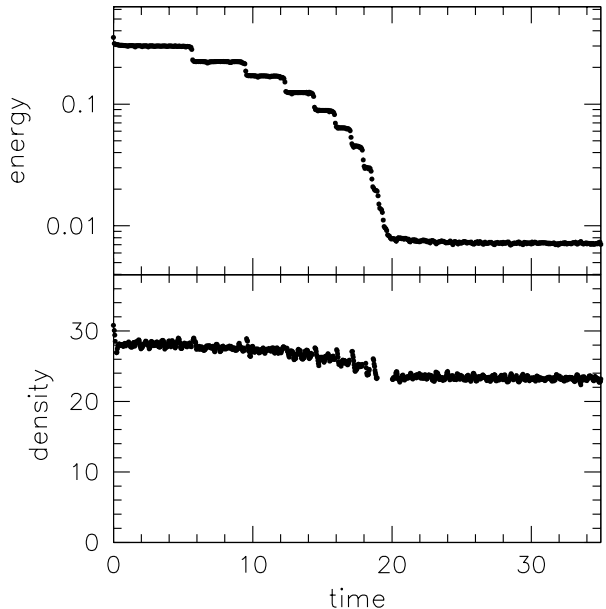}
  \end{center}
  \caption{{\it Upper panel:} Specific energy of the BH particle 
    versus  time in Model A1 with $\vk=0.9\ve$.
    Almost all of the energy loss occurs during passages through the core.
    {\it Lower panel:} Mean density in a sphere of radius 0.05 centered on
    the point of maximum density in the core of the galaxy 
    (excluding the BH).}
  \label{fig:eoft}
\end{figure}
Due to the combined effect of the kick and dynamical friction, the BH
displays a damped oscillatory motion.  The number of radial
oscillations increases with $\vk$; for $\vk=0.9\,\ve$ the BH
experiences $\sim 5$ full radial oscillations before returning to the
core.  Almost all of the energy loss to dynamical friction takes place
during the short intervals that the BH passes through the core.  This
is shown in Figure~\ref{fig:eoft} which plots the evolution of the BH
specific energy $E$ in Model A1 with $\vk=0.9\ve$, where 
\begin{equation} E \equiv \frac{V^2}{2} - \sum_{i=2}^N 
\frac{m_i}{\sqrt{(\mathbf{x}_i-\mathbf{X})^2 + \epsilon^2}}
\label{eq:defe}
\end{equation} 
and the summation is over the ``star'' particles\footnote{Unless
otherwise noted, upper-case variables $X$ and $V$ refer to the BH
particle while lower-case symbols are reserved for the star
particles.}.  The energy lost during the initial emergence from the
core appears to be less than during subsequent passages, suggesting
that dynamical friction requires a finite time to ``turn on'' after
the kick.  During the first few oscillations, the BH's motion remains
essentially rectilinear, but eventually the $Y$- and $Z$-components of
the motion become important due to non-sphericities in the galaxy
potential and also to perturbations from stars.  At late times, the
BH's motion is essentially random, similar to that of a Brownian
particle in a fluid.  Figure~\ref{fig:eoft} also shows the mean
density in a sphere of fixed radius whose center is located at the
estimated density peak (computed via the algorithm described in
\S\,\ref{sec:phaseI}).  The core density decreases rapidly following
the initial ejection, then more gradually as the BH returns again and
again to the core, losing energy to the stars each time.

\begin{figure}
  \begin{center}
    \includegraphics[width=8.5cm]{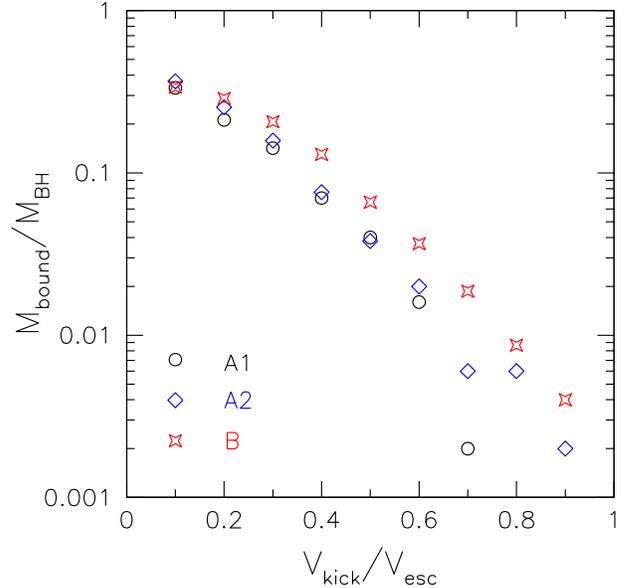}
  \end{center}
  \caption{Stellar mass bound to the BH in the initial models.}
  \label{fig:bound}
\end{figure}
Figure~\ref{fig:bound} shows the mass in stars bound to the BH at
$t=0$.  The bound mass was computed by counting all the stars, within
the influence radius $r_h$, which formed a bound two-body system with
the BH particle.  The influence radius was defined as the radius
containing a mass in stars equal to twice $\mbh$.  The bound mass
decreases steeply with $\vk$, as noted in earlier studies
\citep{MMFHH:04,Boylan:04}, and is ignorable for $\vk\gtrsim 0.6\ve$.

In all cases where the kick velocity was large enough to remove the BH
completely from the core (i.e. $\vk\gtrsim 0.3 \ve$), we observed
three distinct regimes of the motion.  In Phase I, the BH's motion is
well predicted by Chandrasekhar's dynamical friction theory, after
taking into account the changing size of the galaxy core where most of
the friction occurs.  This is the phase illustrated in
Figure~\ref{fig:roft}; in Figure~\ref{fig:eoft}, Phase I extends until
$t\approx 20$.  Phase II begins roughly when the amplitude of the BH's
motion had decayed to the size of the core.  In this phase, the energy
of the BH's orbit continues to decay but with a much longer time
constant than predicted by Chandrasekhar's formula.  The BH and the
core oscillate about their common center of mass in this regime.  In
Phase III, the BH's energy has dropped to the thermal level.  Phase II
is generally longer than Phase I, and this would presumably be even
more true in real galaxies since the amplitude of thermal oscillations
is much lower than in our simulations implying a longer time to reach
the Brownian regime.  We discuss these three regimes in detail below.

\subsection {Phase I}
\label{sec:phaseI}

The extent of Phase I is clearly indicated in the plots of BH energy
vs. time (e.g. Fig.~\ref{fig:eoft}): a distinct ``knee'' appears in
the $E(t)$ curves marking the end of this phase.  Values of $T_I$, the
elapsed time from the kick until the end of Phase I, are given in
Table~\ref{tab:times}.

We compared the evolution of the BH's motion in Phase I with the
predictions of Chandrasekhar's dynamical friction theory
\citep{1943ApJ....97..255C}.  Such comparisons are problematic since
much of the energy exchange between BH and stars occurs during
passages through the galaxy's core, and the core density changes
significantly with time due to the BH's motion.  We dealt with this
problem by breaking the BH's motion into segments, each containing one
passage through the center, and assuming that the galaxy's density
remained constant during each segment.

Chandrasekhar (1943) derived his expression for the dynamical friction
acceleration $F_{\rm df}$ assuming an infinite, homogeneous and
unchanging background of perturbers (stars).  
In the limit that the mass of
the heavy object greatly exceeds the masses of the stars, 
the acceleration is predicted to be 
\begin{equation} F_{\rm df} \approx -2 \pi G^2 \rho\,
\mbh \, \ln(1+\Lambda^2)\, \vbh^{-2} N(<\vbh,r),
\label{eq:df}
\end{equation}
where $\rho(\mathbf{r})$ is the mass density of stars at the BH's 
position, 
$(1/2)\ln(1+\Lambda^2)$ is the Coulomb logarithm, 
$\vbh$ is the BH's instantaneous velocity, 
and  $N(<\vbh,{\mathbf r})$ is the fraction of stars at 
$\mathbf{r}$ that are moving (in the frame of the galaxy) with
velocities less than $\vbh$.

Some care must be taken in the definition of the Coulomb logarithm.
One commonly writes
\begin{equation}
\ln\left(1+\Lambda^2\right) \approx 2\ln\Lambda \approx 2\ln (p_{\rm max}/p_{\rm min})
\end{equation}
where $p_{\rm min}$ and $p_{\rm max}$ are the minimum and maximum
effective impact parameters of the stars that contribute to the
frictional force, and $p_{\rm max}\gg p_{\rm min}$.
However, $p_{\rm min}$ depends on the field-star velocity
\citep{White:49,Merritt:01}
and $p_{\rm max}$ is likewise ill-defined
since a realistic stellar system is inhomogeneous and 
has no outer boundary.

Numerous $N$-body simulations have been carried out to evaluate
Chandrasekhar's formula in the case of a massive particle inspiraling
toward the center of a galaxy
\citep{White:83,Bontekoe:87,Bontekoe:88,Weinberg:89,Cora:97,Bertin:03}.
Early work was typically based on approximate $N$-body schemes and the
results were often discrepant from study to study \citep{Zaritsky:88}.
These differences appear to have been resolved in the last few years
through the use of direct-summation codes
\citep{Spinnato:03,Merritt:06}, which consistently find
$4\lesssim\ln\Lambda\lesssim 6$ for inspiral of massive point
particles, on circular or near-circular orbits, into the centers of
galaxies with steeply-rising density profiles.
Fewer experiments have been done with highly eccentric orbits,
although \cite{Just:05}, using an approximate method, 
find $2\lesssim\ln\Lambda\lesssim 3$ for orbits with moderate eccentricities.

In general, we expect the effective value of $\ln\Lambda$ to be smaller 
for radial orbits than for circular motion.
The dynamical friction force arises from a polarization of the stellar
density which produces an over-dense region, or wake, behind the
massive object \citep{Mulder:83}.  A finite time, of order a galaxy
crossing time, is presumably required for this wake to be set up.  In
the case of a gradually-decaying circular orbit, the galaxy is able to
reach a quasi-steady state after a few orbits of the massive object.
In our case, the position and velocity of the BH are changing
dramatically over one crossing time, so that the wake never has a
chance to establish its steady-state amplitude; indeed just after
apocenter passages, the over-dense region can be seen to lie in {\it
front} of the BH.

In order to determine the effective value of $\ln\Lambda$ in the
$N$-body integrations, we computed BH trajectories using
Chandrasekhar's formula (equation\,\ref{eq:df}) with various values of
the $\ln\left(1+\Lambda^2\right)$ term (henceforth written simply as
$2\ln\Lambda$) and compared them with the $N$-body trajectories.  The
following procedure was followed.

1. The density center of the galaxy moves slightly with respect to the
origin of the coordinates due to transfer of momentum from the kicked
BH to the galaxy. In order to accurately determine the distance of the
BH from the galaxy center as a function of time, we recorded full
snapshots of the particle positions at frequent intervals, then used
the Casertano-Hut (1985) algorithm to find the density center of the
stars in each snapshot.  A smoothing spline was fit through the
measured positions to give a continuous estimate of the center
displacement as a function of time, and this displacement was
subtracted from the BH positions.  (The instantaneous velocity of the
density center was ignored, which is a good approximation at least
until the end of Phase I.)  The resulting correction was at most $\sim
0.02$; at late times the displacement reached a constant value since
the center-of-mass velocity of the system was zero by construction.

2. In order to apply Chandrasekhar's formula we needed to specify the
galaxy model.  The galaxy's mass distribution changes with time due to
the BH's motion; most of this change takes place in the core just
after the BH passes through. We therefore fixed all the parameters in
equation~(\ref{eq:PS1}) except for the core radius $r_b$.  We
determined the effective value of $r_b$ at the discrete times when the
BH passed through the galaxy center by assuming a flat core
($\gamma=0$) and finding the value of $r_b$ such that the mass
contained within $r_b$ according to equation~(\ref{eq:PS1}), with
$\alpha=2$, was the same as the mass in the $N$-body model in a sphere
of radius $r_b$ centered on the BH.  This procedure was always found
to yield a unique $r_b$ and accurately recovered the known value of
$r_b$ in the initial models.

3. BH trajectories were then computed in a piecewise fashion using
Chandrasekhar's formula, starting from one extremum in the BH
displacement and continuing until the next extremum, using the value
of $r_b$ corresponding to the central passage lying between the two
extrema.  This was repeated for several values of $\ln\Lambda$.  We
used equation~(5) of \cite{Szell:05} to compute $N(<\vbh,r)$ in
equation~(\ref{eq:df}) from the assumed $\rho(r)$.

\begin{figure}
  \begin{center}
    \includegraphics[width=8.5cm]{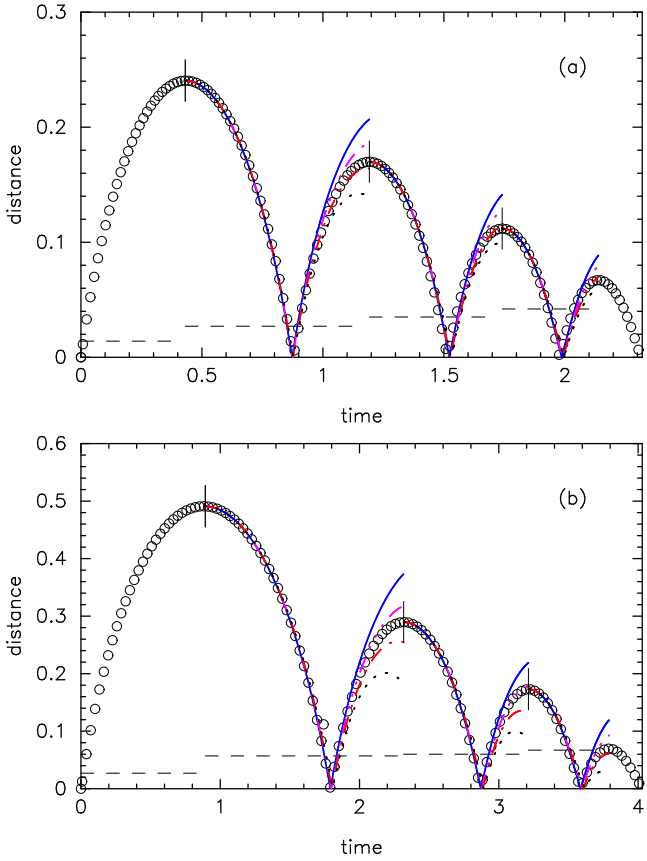}
  \end{center}
  \caption{Comparison between BH trajectories computed via the $N$-body
    integrations (open circles) and via Chandrasekhar's formula
    (\ref{eq:df}) (lines).  The $N$-body models were A1 ($\mbh=0.001$) with
    $\vk=0.7\,\ve$ (a) and B ($\mbh=0.003$) with $\vk=0.8\,\ve$ (b).
    Theoretical trajectories were computed in a piecewise manner,
    starting from extrema in the BH's trajectory (vertical solid lines)
    and continuing until
    the next extremum; the core radius $r_b$ of the galaxy model was
    adjusted as described in the text to give the same core density as
    in the $N$-body model at the time when the BH passed through the
    center.  Horizontal dashed lines show the adopted values of $r_b$.  Line
    colors/styles correspond to different values of $\ln\Lambda$: $1$
    (blue/solid), $2$ (magenta/dashed), $3$ (red/dash-dotted), $4$
    (black/dotted).}
  \label{fig:comp}
\end{figure}

Figure~\ref{fig:comp} shows the results for Model A1 with
$\vk/\ve=0.7$ and Model B with $\vk/\ve=0.8$.  During each inward leg
of the trajectory, the dynamical friction force hardly affects the
motion; only when passing through the dense center is the motion
significantly non-ballistic.  (This could be seen already in
Figures\,\ref{fig:rmax} and~\ref{fig:eoft}.)  The best-fit value of
$\ln\Lambda$ was found to lie in the range
$2\lesssim\ln\Lambda\lesssim 3$, and for such values, Chandrasekhar's
formula did a good job of reproducing the motion.  We found no
evidence of a systematic change in the effective value of $\ln\Lambda$
from one time interval to the next.

\subsection{Phase III}
\label{sec:phaseIII}

The BH trajectories in Figure~\ref{fig:comp} are displayed until the
amplitude of the oscillations has decayed down to roughly the core
radius.  As discussed above, the BH's motion is well predicted by
Chandrasekhar's dynamical friction formula in this regime.  Shortly
after returning to the core, however, the BH's motion was found to
depart strikingly from the predictions of Chandrasekhar's formula.  A
detailed discussion of the motion in ``Phase II'' is presented below.
Before doing so, we consider the motion of the BH at still later
times, ``Phase III,'' when it has reached thermal equilibrium with the
stars.

\begin{figure}
  \begin{center}
 \includegraphics[width=8.5cm]{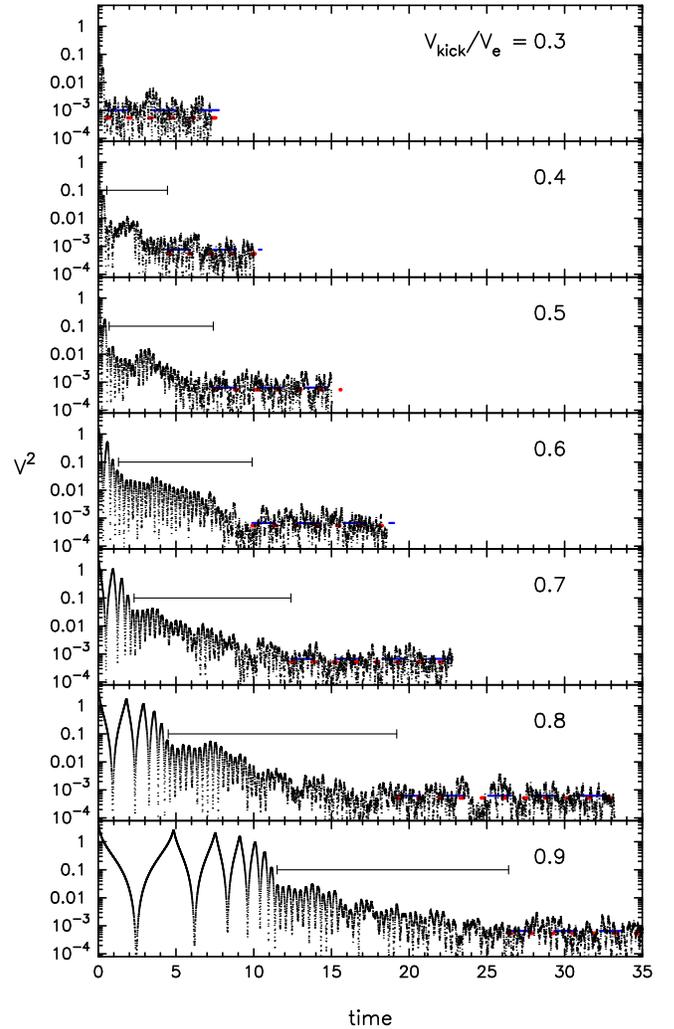}
  \end{center}
  \caption{Squared BH velocity in seven $N$-body integrations of Model
    B.  For $\vk\gtrsim 0.4\ve$ the BH moves completely out of the
    core before falling back.  Ticked, horizontal lines demarcate
    Phase II.  Blue (dashed) lines show $\langle V^2\rangle$ during
    Phase III, and red (dotted) lines show the mean square velocity
    predicted by equation~(\ref{eq:brown}), which assumes that the BH
    particle has reached thermal equilibrium with the stars in its
    vicinity.  }
  \label{fig:voft}
\end{figure}

Figure~\ref{fig:voft} shows the squared velocity of the BH,
$V^2=V_x^2+V_y^2+V_z^2$, over the full integration interval, for kick
velocities $\vk\ge 0.3\ve$ in Model B.  For $\vk\gtrsim 0.4\ve$ the BH
moves substantially beyond the core during its first oscillation
(Fig.~\ref{fig:rmax}).  At late times, the motion of the BH in each of
these integrations appears to be stochastic (i.e. non-quasi-periodic)
but with roughly constant amplitude.

The dashed (blue) lines in this figure show $\left\langle
V^2\right\rangle$, the mean square velocity of the BH averaged over
Phase III.  (The precise definition of the start of Phase III is given
below.)  Also shown (dotted red lines) are estimates of the expected
value of $\left\langle V^2\right\rangle$ for the BH once it reaches
statistical equilibrium with the stars.  The latter velocity, $V_{\rm
Brown}^2$, was computed using 
\begin{equation} V_{\rm Brown}^2 = 3 \frac{m_\star}{\mbh} {\tilde\sigma}^2.
\label{eq:brown}
\end{equation} 
Equation~(\ref{eq:brown}) equates the kinetic energy of the BH
with the mean kinetic energy of a single star in the core.  The
quantity $\tilde\sigma$ is defined as the 1D velocity dispersion of
stars within a sphere of radius $K\times r_h$ centered on the BH, with
$r_h$ the BH's influence radius (the radius containing a mass in stars
equal to twice $\mbh$) and $K$ a constant of order unity.
\cite{MBL:07} used $N$-body simulations to evaluate $K$ for massive
particles at the centers of galaxies with power-law nuclear density
profiles, $\rho\sim r^{-\gamma}$.  They found that $K$ increases
slowly with decreasing $\gamma$, to $K\approx 0.8$ when $\gamma=0.5$.
We set $K=1$ when computing $V_{\rm Brown}$ in Figure~\ref{fig:voft};
the agreement with the measured values is quite good, confirming that
the BH behaves as a Brownian particle in Phase III.

\begin{figure}
  \begin{center}
    \includegraphics[width=8.5cm]{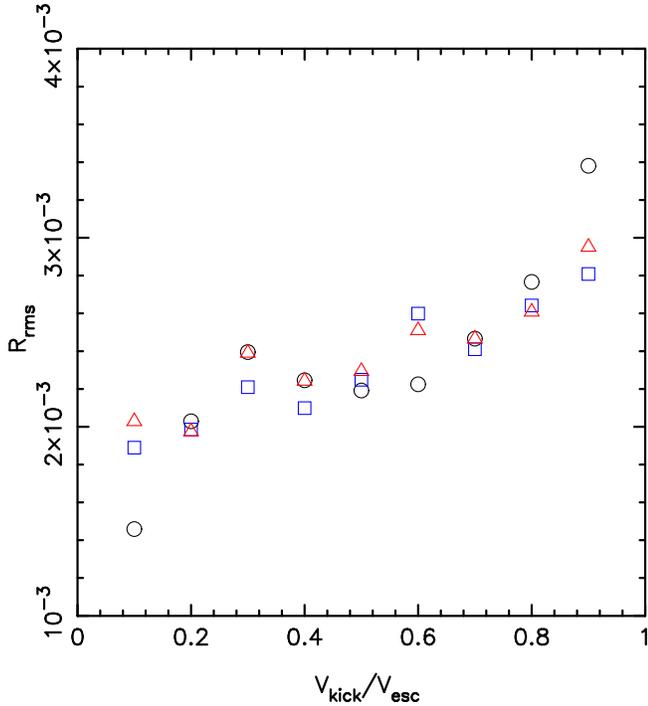}
  \end{center}
  \caption{RMS amplitude of the BH oscillations in the Brownian
    regime, Phase III, for models A1 (black/circles), A2
    (blue/squares) and B (red/triangles).}
  \label{fig:rbrown}
\end{figure}

Figure~\ref{fig:rbrown} shows the rms amplitude of the BH's motion
averaged over Phase III.
Since the density center of the galaxy drifts, as described above,
smoothing splines were first fit to the ${\bf X}(t)$ values
for the BH and the rms deviations were computed with respect
to the smoothed trajectories.
Figure~\ref{fig:rbrown} shows a general trend of increasing $R_{\rm rms}$ with
decreasing core density, as expected if the motion in this
regime obeys the virial theorem,
\begin{equation}
\langle V^2 \rangle \approx \frac{4}{3} \pi G \rho_c \langle R^2\rangle.
\label{eq:rbrown}
\end{equation} 
This relation \citep[cf.][]{BW:76} assumes a constant-density
core, ignores the back-reaction of the BH's motion on the stars, and
ignores any coupling between random gravitational perturbations from
the stars and the quasi-periodic motion of the BH in the smooth
potential of the core.  Nevertheless, equation~(\ref{eq:rbrown}) was
found to reproduce the measured $R_{\rm rms}$ values in
Figure~\ref{fig:rbrown} quite well if $\rho_c$ was defined as the mean
density of stars within $r_h$.  Fluctuations in $R_{\rm rms}$ about
the mean relation in Figure~\ref{fig:rbrown} appear to be due
primarily to fluctuations in $V_{\rm rms}$ and would presumably be
smaller if the $R_{\rm rms}$ values were averaged over longer time
intervals.  The near agreement between the $R_{\rm rms}$ values for
the runs with small and large $\mbh$ is a consequence of the larger
core size / lower core density in runs with larger $\mbh$, which
compensates for the lower $\langle V^2\rangle\propto \mbh^{-1}$.

We note here that the amplitude of the BH's Brownian motion is always
a factor 10 or more smaller than the final core radii of the models
(Table~\ref{tab:fits}).  This implies that the motion of the BH when
it first returns to the core -- at the start of Phase II -- should not
be appreciably affected by discreteness effects, i.e. by perturbations
from individual stars.  This conclusion is confirmed below.

We note also that the amplitude of Brownian oscillations of BHs in
real galaxies (expressed as a fraction of the galaxy effective radius,
say) would be smaller than in our models by the factor $\sim
\sqrt{(M_{\rm gal}/m_\star)/N}$, i.e. $\sim 50$ for $M_{\rm
gal}=10^9M_\odot$ and $\sim 500$ for $M_{\rm gal}=10^{11}M_\odot$.
The time required for a BH to reach these lower kinetic energies would
also presumably be longer than in our simulations, as discussed in
more detail below.

\subsection{Phase II}
\label{sec:phaseII}

As noted above, the motion of the BH after returning to the core, and
before reaching the Brownian regime, is not well described by
Chandrasekhar's formula.  Here we consider the motion in this regime
(``Phase II'').

Figure~\ref{fig:voft} reveals the following qualitative features.

1. The motion in Phase II is essentially oscillatory, with a period
similar to that at the end of Phase I, i.e. roughly equal to the
period of oscillation of a test particle moving in the stellar core.

2. There is evidence of additional frequencies affecting the BH's
motion.  For instance, the amplitude of the oscillations sometimes
appears to {\it increase} temporarily over several periods in a manner
suggestive of beats.

3. Averaged over many periods, the mean amplitude of the oscillations
decays, but with a time constant that is much longer than observed
toward the end of Phase I.

4. Near the end of Phase II, the motion becomes increasingly
stochastic, presumably due to perturbations from individual stars.
Eventually the BH rms velocity falls to the Brownian (thermal) level
marking the start of Phase III.

5. Phase II always begins roughly when the stellar mass interior to
the BH's orbit is equal to $\mbh$.  When $\vk\lesssim 0.3\ve$, the BH
never escapes the core, and its motion appears to transition directly
from Phase I to Phase III.

Based on Figure~\ref{fig:voft}, the elapsed time in Phase II can be
substantially longer than the time spent in Phase I.  Understanding
the character of the motion in this regime is therefore crucial for
predicting the expected displacement of a supermassive BH in a real
galaxy following a kick.

We begin by considering a simple model for damped oscillations of a
massive particle in a constant-density core.  While this model will
fail to quantitatively reproduce the motion in Phase II, it provides a
useful framework for discussing what is observed in the simulations.

In the absence of dynamical friction, and neglecting the influence
of the massive particle's presence on core structure,
the motion of the massive particle is simple harmonic oscillation 
with frequency
$\omega_c = \sqrt{(4\pi/3)G \rho_c}$;
$\rho_c$ is the core density, assumed constant within
a radius $r_c$.
To this motion we add the acceleration due to dynamical friction.
If the velocity distribution of the stars that produce the friction
is Maxwellian with 1D velocity dispersion $\sigma_c$, and if
the BH's velocity satisfies $V\ll\sigma_c$, 
the resulting equation of motion in any coordinate $x_i$ is
\begin{equation} \ddot{X_i} + \tdf^{-1}\dot{X_i} + \omega_c^2 X_i = 0
\label{eq:df1}
\end{equation}
where 
\begin{equation}
\tdf = \frac{3}{8} \sqrt{\frac{2}{\pi}} \frac{\sigma_c^3}{ G^2\rho_c\mbh\ln\Lambda}
\end{equation}
is the dynamical friction damping time \citep{Merritt:85}.
The condition for underdamped oscillations is $2\omega_c\tdf> 1$,
where
\begin{mathletters}
  \begin{eqnarray}
    2\omega_c\tdf &=& \frac{\sqrt{6}}{2} 
    \frac{\sigma_c^3}{G^{3/2}\rho_c^{1/2}\mbh\ln\Lambda}
    \label{eq:damp} \\
    &=& \frac{\sqrt{6\pi}}{9} \form^3 
    \frac{M_c}{\mbh\ln\Lambda},
  \end{eqnarray}
\end{mathletters}
with $M_c \equiv (4/3) \pi \rho_cr_c^3$ the core mass;
the second relation uses the ``core-fitting'' formula
of \citet{RPKK:72},
\begin{equation} 
\sigma_c^2 = \form^2 \frac{4\pi}{9}G\rho_c r_c^2.
\label{eq:King}
\end{equation} 
$\form \approx 2$ for our models.
Thus
\begin{equation}
2\omega_c \tdf \approx  4 \frac{M_c}{\mbh\ln\Lambda}
\end{equation}

\begin{figure}
  \vspace*{-1mm}
  \begin{center}
    \includegraphics[width=8.5cm]{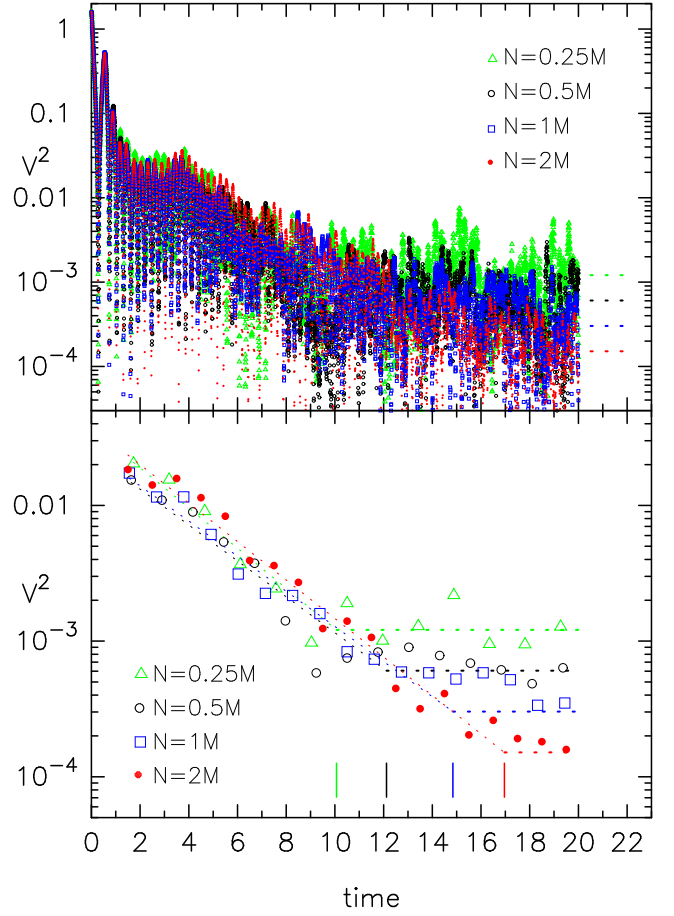}
  \end{center}
  \caption{Evolution of the BH kinetic energy in a series of integrations
    of model B with various $N$, and $\vk=0.6\ve$.
    {\it Top panel:} Squared velocity of the BH versus time.
    Dashed lines at the right show the predicted values of $V^2$
    in the Brownian regime (eq.~\ref{eq:brown}).
    {\it Bottom panel:} Binned values of $V^2$ in Phases II and III.
    Dotted lines are least-squares fits to the binned data.
    These fits are plotted until the time at which they
    intersect the Brownian $V^2$; these times are marked
    by the vertical solid lines.
    The latter are found to be spaced with roughly constant
    separation indicating that the time required for the BH to reach 
    thermal equilibrium with the stars increases roughly as $\ln N$.}
  \label{fig:tests}
\end{figure}

In our simulations (and in real galaxies), the right hand side of this
expression is $\gtrsim 1$, since core masses are $\sim $ a few $\mbh$
\citep{Merritt:06} and $2\lesssim \ln\Lambda\lesssim 3$
(\S\,\ref{sec:phaseI}).  It follows that the motion of the BH should
be under-damped, though not far from critically damped, after it
re-enters the core.  The solutions to equation (\ref{eq:df1}) in the
under-damped regime are
\begin{equation} 
X_i(t) = A_i\,e^{-t/2\tdf} \sin\left(\omega_c t + \phi_i\right).
\label{eq:xioft}
\end{equation} 
Writing $\Theta\equiv 2\omega_cT_{\rm df}\gtrsim 1$ and
$T_c\equiv 2\pi/\omega_c$, the energy decay time is predicted to be
$T_{\rm df}=(\Theta/4\pi)T_c$, i.e. shorter than the orbital period.
Such short decay times are in fact observed near the end of Phase I
(Figure~\ref{fig:comp}).

However, Figure~\ref{fig:voft} shows that this is not the case in
Phase II: the mean damping time is substantially longer than an
orbital period.  The abrupt decrease in the energy dissipation rate at
the start of Phase II can also be seen in Figure~\ref{fig:eoft}(a).

A possible explanation for the slower damping in Phase II is
discreteness effects: perturbations from individual stars, some of
which act to accelerate the BH, become increasingly competitive with
mean-field effects (including dynamical friction) as the BH moves more
slowly.  Indeed, in the Brownian regime (Phase III), the accelerating
perturbations are equally as strong, in a time-averaged sense, as
dynamical friction.  While the amplitude of the BH oscillations at the
onset of Phase II is always much greater than the Brownian amplitude
in these simulations (cf.~Fig.~\ref{fig:rbrown} and the accompanying
discussion), it is still conceivable that discreteness effects are
responsible for the anomalously slow decay of the BH's orbit at this
time.

To securely rule out this possibility, we repeated the integration of
model B with $\vk=0.6\ve$, increasing $N$ up to $N=2\times 10^6$.
Figure~\ref{fig:tests} shows the results.  The slowly-damped
oscillations in Phase II are clearly not an artifact of a too-small
$N$.  In all cases, for instance, the fifth extremum in $V^2$ (which
occurs at $t\approx 1.42$) is comparable or greater in amplitude to
the fourth extremum (at $t\approx 1.16$), rather than being much lower
in amplitude as would be expected from the above analysis or from
Figure~\ref{fig:comp}.  We also carried out a number of tests varying
the integration time-step parameter $\eta$; again, no systematic
dependence of the evolution in Phase II on this parameter was
observed.

Particularly striking in Figure~\ref{fig:tests} is the accurately
exponential decay of the BH's kinetic energy throughout Phase II; this
is clearest in the simulation with largest $N$, where the exponential
damping continues over two decades in energy.  We note again that an
exponentially decaying energy is predicted by the simple model just
presented, but the model predicts a much shorter time constant than
what is observed in the $N$-body simulations.

\begin{figure}
  \begin{center}
    \includegraphics[width=8.5cm]{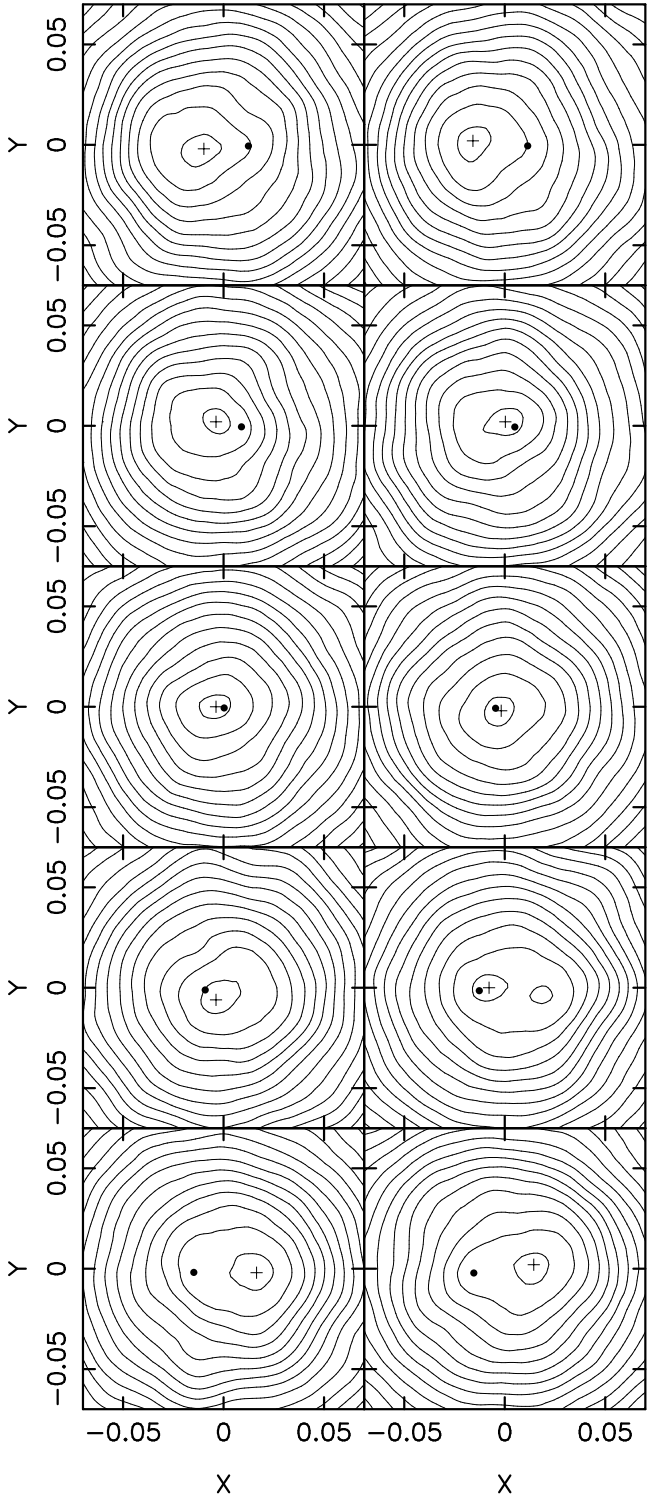}
  \end{center}
  \caption{Core-BH oscillations in Phase II.  This is the $N=2\times
    10^6$ integration of Model B shown as the filled (red) circles in
    Fig.~\ref{fig:tests}.  Contours are separated by $0.034$ in
    $\log_{10}$ of the projected density.  Filled circles mark the BH
    and crosses mark the approximate location of the (projected)
    stellar density maximum.  Times are
    $t=2.1875,2.21875,2.25,...2.46875$, increasing from upper left to
    lower right.  The elapsed time in this figure spans approximately
    1/2 oscillation period of the BH.  }
  \label{fig:cont}
\end{figure}

Figure~\ref{fig:cont} suggests why Chandrasekhar's (1943) formula
might break down in Phase II.  The approximation of a stationary
galaxy is strongly violated in this regime.  The galaxy's density
center oscillates with opposite phase to the BH, and with roughly the
same frequency and amplitude.  This is consistent with the observation
that Phase II always begins roughly when the mass in stars inside the
BH's orbit is similar to $\mbh$.  Evidently, in this regime, the BH
and the core oscillate about their common center of mass as a
two-body system.  Chandrasekhar's derivation, which assumed a body on
a linear trajectory through an infinite homogeneous medium, is
unlikely to apply to oscillations like those in Figure~\ref{fig:cont},
since the BH is periodically accelerated, then decelerated, by the
density peak.  The rate at which such oscillations decay is known to
be sensitively dependent on resonant interactions \citep{TW:84} and
can be arbitrarily low \citep{Louis:88,Sridhar:89,SN:89,MFR:90},
although we are not aware of any theoretical treatment that is
directly applicable to oscillations like those in
Figure~\ref{fig:cont}.

Ours is not the first $N$-body study to observe persistent
oscillations of massive objects at the centers of $N$-body models.
\cite{Miller:92} and \cite{Miller:96} reported a series of $N$-body
integrations, using a grid-based code, of a disk at the center of an
axisymmetric galaxy model.  They observed what appeared to be
over-stable oscillations of particles initially at rest near the
center of the disk; the oscillation frequency was roughly
$\sqrt{(4\pi/3)G\rho_c}$ and the maximum amplitude was roughly the
size of the core.  All of these features are characteristic of the
oscillations that we observe in Phase II.  \cite{Miller:92} also
reported ``a couple of experiments in which a massive object was put
into orbit within a galaxy model,'' presumably near the center, and
observed ``residual oscillations'' with amplitude roughly equal to the
radius at which the enclosed mass was equal to the object's mass,
again similar to what we observe.  \cite{Miller:92} briefly describe a
model for the oscillations, in which periodic motion of the core as a
whole, at roughly the same frequency as core internal frequencies,
drives the oscillations.

A number of other $N$-body studies have noted a decrease in the
effective value of $\ln\Lambda$ once a massive object has spiraled
into a constant-density core.  Typically the observed decrease is
modest, a factor$\sim 2-3$ or so
\citep{Bontekoe:87,Bontekoe:88,Weinberg:89,Cora:97}, although one
recent study \citep{Read:06} found a nearly complete disappearance of
dynamical friction after the infalling particle reached the core. Read
et al. proposed that the apparent vanishing of the dynamical friction
force in their simulations could be explained by the degeneracy of
orbital frequencies in the harmonic-oscillator potential corresponding
to a precisely flat, central density profile. In such models, Read et
al. found that the disappearance of the dynamical friction force was
critically dependent on whether the plane defined by the inspiralling
particle's orbit remained fixed; precession, induced e.g. by
finite-$N$ perturbations, caused dynamical friction to turn on again,
though at a rate much slower than expected from Chandrasekhar
friction.  While Read et al. did consider the effect of varying the
initial log-slope of the background distribution, their models were
always spherical.  The cores in our models are not precisely flat nor
are our models precisely spherical (once the BH particle has been
ejected) and these differences (coupled with the fact that the
gravitational potential of the core is highly oscillatory in Phase II)
may explain why we do not observe the dramatic stalling reported by
Read et al.  In any case, the apparent lack of an $N$-dependence in
our simulations (Figure~\ref{fig:tests}) suggests that the critical
difference between our results and those of Read et al. is not
particle number.

The Phase II oscillations were clearly visible in every integration
with $\vk\ge 0.4\ve$.  For $\vk=0.3\ve$ there were hints of a delayed
return to the Brownian regime in some of the integrations
(e.g. Fig.~\ref{fig:voft}) but not to the extent that we were able to
estimate damping times.  We could not detect the Phase II oscillations at all
for $\vk\le 0.2\ve$; in these integrations, the BH kinetic energy
appears to drop very rapidly after the return to the core, more or
less as expected based on the analytic model presented above or by an
extrapolation of the behavior in Phase I.  In any case, we assume in
the remainder of this paper that the Phase II oscillations are absent
when $\vk\le 0.3\ve$.  Integrations with much larger $N$ might modify
this conclusion.

The occasional {\it increase} in the amplitude of the Phase II
oscillations, which is seen in virtually all the integrations, is
suggestive of a dynamical instability \citep{Tremaine:05}.  However an
instability would presumably act even in the case of small kicks,
while as noted above, Phase II oscillations appear to be absent for
$\vk\lesssim 0.3\ve$.  We speculate that the BH must be kicked
completely out of the core in order for the BH-core oscillations to be
excited, as suggested by \cite{Miller:92}.  The roughly sinusoidal
variations in the envelope of $V^2(t)$, with a much lower frequency
than $\omega_c$, could naturally be explained in terms of beating,
e.g. between the frequency of motion in the core and the frequency at
which the core itself oscillates in the galactic potential.

Figures~\ref{fig:voft} and~\ref{fig:tests} suggest that the core-BH
oscillations in Phase II decay roughly as an exponential in time, at
least when viewed through a window of several orbital periods or
longer.  We investigated a number of ways to quantify the time
constant $\tau$ associated with the energy damping:

1. Plots of BH energy versus time (equation~\ref{eq:defe},
  Fig.~\ref{fig:eoft}) were found not to be very useful in this regard
  since the total energy is dominated by the potential energy which
  exhibits fairly large fluctuations from time step to time step.

2. In a constant-density core, the unperturbed motion is simple
harmonic oscillation with frequency $\omega$ and energy
$$
E_{\rm SHO} = \frac{1}{2} \sum_{i=1}^3 \left(\omega^2 X_i^2 + V_i^2\right).
$$ 
We determined the dominant frequency of the BH's motion in Phase II
by carrying out discrete Fourier transforms of the complex functions
$X_i(t)+iV_i(t)$ and constructing power spectra \citep[e.g.][]{Laskar:90}.
Least-squares fits to $\ln E_{\rm SHO}$ vs. $t$ were then carried out
to find the damping time constant.  This approach was reasonably
objective and robust, but can be criticized on the grounds that the
core density is not constant and the density center is moving with time
(Fig.~\ref{fig:cont}), making the interpretation of $E_{\rm SHO}$
problematic.

\begin{figure}
  \begin{center}
    \includegraphics[width=8.5cm]{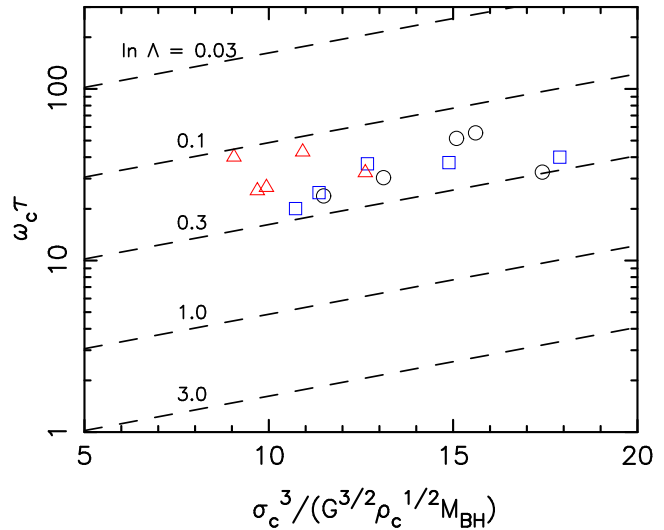}
  \end{center}
  \caption{Energy-decay time constants $\tau$ for the BH in Phase II,
    for models A1 (black/circles), A2 (blue/squares) and B (red/triangles).
  }
  \label{fig:tauomega}
\end{figure}

3. Given the difficulties with evaluating and interpreting the total
energy of the BH, we chose in the end to quantify the energy damping
purely in terms of the the BH's kinetic energy.  As noted above
(Fig.~\ref{fig:tests} and associated text), $V^2(t)$ exhibits a nicely
exponential decay with a well-defined time constant, and the decay is
observed to continue over $\sim 2$ decades in kinetic energy in the
case of the simulation with the largest $N$, until the BH's kinetic
energy reaches the Brownian value.  We evaluated the associated time
constant by carrying out least-squares fits of $\ln V^2$ to time,
yielding the coefficients $(V_I^2, \tau)$ in the expression 
\begin{equation}
V^2(t) \approx V_I^2\,e^{-\left(t-T_I\right)/\tau}.
\label{eq:vfit}
\end{equation}
Table~\ref{tab:times} gives the $\tau$ values derived from this method.
We present results only from $N$-body integrations with $\vk\ge 0.4\ve$
since the smaller kicks did not excite distinct BH-core oscillations,
as discussed above.
To the extent that the motion approximates a
damped SHO, the energy damping time is identical
to the time constant for decay of the kinetic energy alone,
and henceforth we will refer to $\tau$ as the ``energy damping
time constant.''
However in practice, we will use equation~(\ref{eq:vfit})
only to predict changes in $\langle V^2\rangle$.

The energy damping times in Table~\ref{tab:times} can immediately be
scaled to physical units using equation~(\ref{eq:units}).  Such a
scaling presumes that the core properties of our $N$-body models --
which presumably determine $\tau$ -- are related to global properties
in the same way as in real galaxies.  A better scheme would relate
$\tau$ directly to the parameters $(\rho_c,\sigma_c,\mbh)$ that
describe the conditions in the core.  Since we do not understand the
mechanism(s) responsible for the orbital damping in Phase II, we
experimented with several ways of plotting $\tau$ versus core
parameters.

Figure~\ref{fig:tauomega} shows that a reasonably tight correlation
exists when $\omega_c\tau$ is plotted against
$\sigma_c^3/(G^{3/2}\rho_c^{1/2}\mbh)$.  This is the expected
dependence if dynamical friction is responsible for the damping
(cf. equation~\ref{eq:damp}).  However, the effective value of
$\ln\Lambda$ needed to produce the measured damping times is very
small, $0.1 \lesssim\ln\Lambda\lesssim 0.3$
(Figure~\ref{fig:tauomega}).  This is yet another way of stating that
orbital decay in Phase II is much slower than predicted by
Chandrasekhar's formula -- roughly a factor $10-20$, if we adopt
$\ln\Lambda\approx 2.5$ for the expected value of the Coulomb
parameter (Fig.~\ref{fig:comp}).

\begin{table}
\begin{center}
\caption{Times associated with the evolution in Phases I and II}
\label{tab:times}
\begin{tabular}{cccccccc}
\hline
$\vk/\ve$ & $T_I$ &  $\tau$ & $T_{II}$ & &$T_{II}$, &$N_{\rm gal}=$ ... & \\
          &       &         &          &$3\times 10^9$&$3\times10^{10}$&$3\times10^{11}$  & $3\times10^{12}$\\
\hline
\multicolumn{8}{c}{A1}\\
\hline
0.1 & 0.3 & --  & --   & --   & --   & --   & --   \\
0.2 & 0.3 & --  & --   & --   & --   & --   & --   \\
0.3 & 0.3 & --  & --   & --   & --   & --   & --   \\
0.4 & 0.4 & 1.6 & 2.9  & 16.8 & 20.5 & 24.2 & 27.9 \\
0.5 & 0.7 & 1.3 & 3.2  & 14.5 & 17.5 & 20.5 & 23.4 \\
0.6 & 1.5 & 1.9 & 4.6  & 21.1 & 25.5 & 30.0 & 34.2 \\
0.7 & 3.0 & 3.4 & 5.8  & 35.3 & 43.2 & 51.0 & 58.8 \\
0.8 & 7.3 & 3.8 & 8.5  & 41.6 & 50.3 & 59.0 & 67.8 \\
0.9 & 20.2 & 2.5 & 6.5 & 28.3 & 34.0 & 39.8 & 45.5 \\
\hline
\multicolumn{8}{c}{A2}\\
\hline
0.1 & 0.3 & --   & --   & --   & --   & --   & --   \\
0.2 & 0.3 & --   & --   & --   & --   & --   & --   \\
0.3 & 0.3 & --   & --   & --   & --   & --   & --   \\
0.4 & 0.4 & 1.0  & 1.5  & 10.2 & 12.5 & 14.8 & 17.1 \\
0.5 & 0.7 & 0.95 & 1.9  & 10.2 & 12.4 & 14.5 & 16.7 \\
0.6 & 1.3 & 1.3  & 3.4  & 14.7 & 17.7 & 20.6 & 23.6 \\
0.7 & 2.7 & 2.1  & 4.4  & 22.7 & 27.5 & 32.3 & 37.2 \\
0.8 & 6.5 & 2.4  & 4.6  & 25.4 & 31.0 & 36.5 & 42.0 \\
0.9 & 20.0 & 2.8 & 6.7  & 31.1 & 37.5 & 43.9 & 50.4 \\
\hline
\multicolumn{8}{c}{B}\\
\hline
0.1 & 0.5 & --   & --   & --   & --   & --   & --   \\
0.2 & 0.5 & --   & --   & --   & --   & --   & --   \\
0.3 & 0.5 & --   & --   & --   & --   & --   & --   \\
0.4 & 0.55& 2.2  & 3.9  & 23.0 & 28.1 & 33.2 & 38.2 \\
0.5 & 0.7 & 2.8  & 6.7  & 31.1 & 37.5 & 43.9 & 50.4 \\ 
0.6 & 1.3 & 2.6  & 10.8  & 33.4 & 39.4 & 45.3 & 51.4 \\
0.7 & 2.3 & 2.9  & 10.1 & 35.3 & 42.0 & 48.7 & 55.3 \\
0.8 & 4.5 & 5.2  & 14.7 & 60.0 & 71.9 & 83.9 & 95.8 \\
0.9 & 11.5 & 4.3 & 14.9 & 52.3 & 62.2 & 72.1 & 82.0 \\
\hline
\end{tabular}
\end{center}
\end{table}

In terms of this scaling, Figure~\ref{fig:tauomega}
allows us to express the damping times in Phase II as
\begin{mathletters}
\begin{eqnarray}
\tau &\approx & 15 \frac{\sigma_c^3}{G^2 \rho_c \mbh}\\
&\approx& 3\times 10^7 {\rm yr} 
\left(\frac{\sigma_c}{200\kms}\right)^{-3.86}
\left(\frac{r_c} {30\,{\rm pc}}\right)^2
\label{eq:tauomega}
\end{eqnarray}
\end{mathletters}
where the second line uses the $\mbh-\sigma$ relation \citep{FF:05}.
Based on Figure~\ref{fig:tests} and on the other
arguments given above, we expect the scaling
in equation~(\ref{eq:tauomega}) to be independent of $N$,
i.e. of stellar mass.

Table~\ref{tab:times} also gives estimates of $T_{II}$, the elapsed
time in Phase II.  We defined $T_{II}$ as the time, measured from the
end of Phase I, required for the BH's velocity to fall to its rms
value in Phase III, assuming the time dependence of
equation~(\ref{eq:vfit}).  Table~\ref{tab:times} shows that $T_{II}$
is typically longer than $T_I$.

In a galaxy with $\gg 10^6$ stars, $V_{\rm Brown}^2$ would be much
lower, and $T_{II}$ correspondingly longer, than in our models.
Assuming that the exponential dependence of energy on time persists to
arbitrarily low values of $E$, the additional time spent in Phase II
would be 
\begin{equation} 
\tau\ln\left(N_{\rm gal}/N\right)
\label{eq:addtime}
\end{equation}
where $N_{\rm gal}$ is the number of stars in the galaxy.
We used the set of $N$-body simulations in Figure~\ref{fig:tests}
to test this dependence.
According to equation~(\ref{eq:addtime}), doubling the number 
of particles should  extend the elapsed time in Phase II
by an additive amount of $\tau\ln 2 = 0.693\tau\approx 2.2$,
given that the mean $\tau$ value for the four integrations
is $3.2$.
Figure~\ref{fig:tests} confirms this prediction for
$0.25\times 10^6\le N \le 2\times 10^6$.

Accordingly, Table~\ref{tab:times} also gives values of $T_{II}$
calculated from this formula, for $N_{\rm gal}=(3\times 10^9, 3\times
10^{10}, 3\times 10^{11}, 3\times 10^{12})$.  Conversion from the
$N$-body units of Table~\ref{tab:times} to years is discussed in \S\,\ref{sec:obs}.

The exponential nature of the damping implies that the distribution 
of displacements during Phase II is
approximately uniform in $\ln\Delta r$.

\section{Effects on the stellar distribution}
\label{sec:profiles}

The displacement of the BH due to gravitational radiation recoil
affects the stellar distribution and therefore the density profile of
the host galaxy. We expect the stellar structure inside the core to be
particularly affected by the motion of the BH, with important
implications for the shape of the brightness profile in the inner
region.  In order to evaluate the changes induced by the escaping BH,
we constructed spatial and projected density profiles for all models
at the end of the simulations, when the BH is well into the Brownian
regime.

\begin{figure*}
  \begin{center}
    \includegraphics[width=17cm]{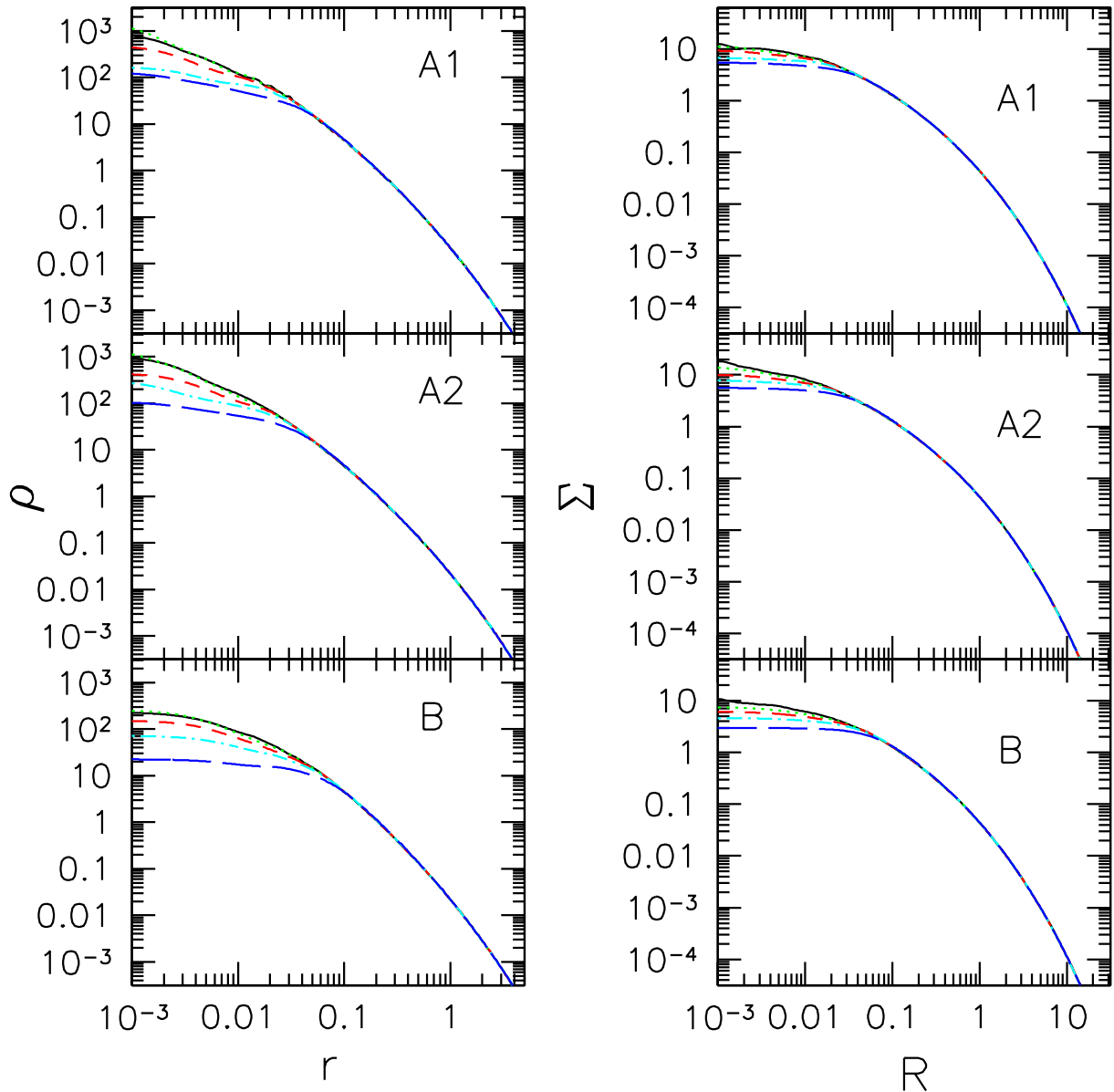}
  \end{center}
  \caption{Space (left) and projected (right) density profiles
    for models A1, A2 and B and different values of the kick
    velocity: $\vk = 0.1\,\ve$ (green/dotted), $\vk = 0.3\,\ve$
    (red/dashed), $\vk = 0.5\,\ve$ (cyan/dot-dashed), $\vk = 0.9\,\ve$
    (blue/long dashed).  The black solid lines represent the initial
    profile, which is the same for each value of $\vk$.}
  \label{fig:density}
\end{figure*}

Figure~\ref{fig:density} shows the space (left plots) and projected
(right plots) density profiles in models A1, A2, B for $\vk =
(0.1,0.3,0.5,0.9)\,\ve$.  We constructed these density profiles using
the kernel-based algorithm of \cite{Merritt:06b}.  Particle positions
were first shifted to coordinates that placed the BH at the origin.
The algorithm uses an angle-averaged Gaussian
kernel and modifies the kernel width based on a pilot
(nearest-neighbor) estimate of the density in order to maintain a
roughly constant ratio of bias to variance in the final density
profile.  The projected density $\Sigma(R)$ was computed via numerical
projection of the space density.  In order to reduce the noise still
further, we combined multiple snapshots at late times and performed
the fit on the combined data sets.

Figure~\ref{fig:density} shows that a large core develops in the
simulations due to the escape of the BH and its several passages
through the central region. As the BH oscillates under the effect of
the kick, it transfers energy to the surrounding stars, thus pushing
them to larger distances.  The stellar density in the core drops and
the slope of the inner distribution decreases, leaving an inner
profile that is flatter than the initial one. The amount of flattening
in the profile or, equivalently, the mass deficit with respect to the
initial profile (shown in the figure with the black solid lines),
increases monotonically with the kick velocity.

It is interesting to assess whether the final $N$-body profiles are 
consistent with the core-S{\'e}rsic law, which is commonly fit
to galaxies with evacuated cores \citep{Graham:03}.
The core-S{\'e}rsic law is:
\begin{mathletters}
\begin{eqnarray}
\Sigma(R) & = &
\Sigma^{'}\left[1+\left(\frac{\rb}{R}\right)^{\alpha}\right]^{\gamma/\alpha}e^{-b
\left[\left(R^{\alpha}+\rb^{\alpha}\right)/\re^{\alpha}\right]^{1/n\alpha}},\\
\Sigma^{'} & = & \Sigma_b \, 2^{-\gamma/\alpha} \, e^{b
\left(2^{1/\alpha}\,\rb/\re \right)^{1/n}}\, , 
\end{eqnarray}
\end{mathletters}
where 
$\Sigma_b$ is the density at the break radius $r_b$ and the other parameters are as in 
equations~(\ref{eq:PS1}) and~(\ref{eq:PS2}).  
To carry out the fits in a manner as similar as possible to
the procedure followed by observers, we counted the projected 
particle positions in bins equally spaced in $\log R$.
The parameters $(\re,\rb,\alpha,n,\Sigma^{'})$ were then
varied until the summed residuals in $\mu=-2.5\log\Sigma$ 
were minimized.

\begin{table}
\begin{center}
\caption{Fit parameters for models A1, A2 and B.}
\label{tab:fits}
\begin{tabular}{c|rrrrrr}
$\vk$ & $\rb$ &  $n$ & $\alpha$ & $\gamma$ & $\re$ & $\Sigma_b$\\
\hline 
\multicolumn{7}{c}{A1}\\
\hline
0.1 & 0.013 & 4.04 & 3.1  &  0.21 & 0.93 & 6.3 \\ 
0.2 & 0.016 & 4.05 & 4.2  &  0.24 & 0.93 & 5.6 \\ 
0.3 & 0.017 & 4.05 & 4.1  &  0.19 & 0.93 & 5.4 \\ 
0.4 & 0.018 & 4.06 & 3.7  &  0.14 & 0.93 & 5.1 \\ 
0.5 & 0.019 & 4.06 & 3.6  &  0.11 & 0.93 & 4.9 \\ 
0.6 & 0.020 & 4.06 & 3.5  &  0.08 & 0.93 & 4.6 \\ 
0.7 & 0.021 & 4.07 & 3.1  &  0.04 & 0.92 & 4.3 \\ 
0.8 & 0.022 & 4.07 & 2.8  &  0.02 & 0.92 & 4.1 \\ 
0.9 & 0.024 & 4.05 & 2.9  &  0.05 & 0.93 & 3.9 \\ 
\hline 
\multicolumn{7}{c}{A2}\\
\hline
0.1 &  0.014  & 4.04 & 7.4  &  0.34 & 0.93 &  6.5 \\ 
0.2 &  0.015  & 4.04 & 6.7  &  0.31 & 0.93 &  6.2 \\ 
0.3 &  0.016  & 4.04 & 6.5  &  0.25 & 0.92 &  5.9 \\ 
0.4 &  0.015  & 4.05 & 3.5  &  0.14 & 0.92 &  5.9 \\ 
0.5 &  0.017  & 4.05 & 4.0  &  0.12 & 0.92 &  5.4 \\ 
0.6 &  0.019  & 4.05 & 4.2  &  0.16 & 0.92 &  4.9 \\ 
0.7 &  0.018  & 4.07 & 2.7  &  0.08 & 0.92 &  4.9 \\ 
0.8 &  0.022  & 4.06 & 3.6  &  0.07 & 0.92 &  4.4 \\ 
0.9 &  0.026  & 4.07 & 4.1  &  0.11 & 0.92 &  3.9 \\ 
\hline 
\multicolumn{7}{c}{B}\\
\hline
0.1 &  0.020  & 4.05 & 1.9  &  0.16 & 0.92 & 4.4  \\ 
0.2 &  0.026  & 4.04 & 2.9  &  0.20 & 0.93 & 3.8  \\ 
0.3 &  0.030  & 4.04 & 3.7  &  0.20 & 0.93 & 3.4  \\ 
0.4 &  0.034  & 4.05 & 4.3  &  0.16 & 0.92 & 3.1  \\ 
0.5 &  0.034  & 4.06 & 3.1  &  0.12 & 0.92 & 3.0  \\  
0.6 &  0.035  & 4.06 & 3.0  &  0.09 & 0.92 & 2.8  \\ 
0.7 &  0.039  & 4.08 & 3.0  &  0.07 & 0.91 & 2.6  \\ 
0.8 &  0.042  & 4.08 & 2.9  &  0.05 & 0.91 & 2.4  \\ 
0.9 &  0.044  & 4.09 & 2.6  &  0.02 & 0.91 & 2.3  \\ 
\hline
\end{tabular}
\end{center}
\end{table}

The best-fit parameters for models A1, A2 and B are listed in
Table~\ref{tab:fits}.
\begin{figure}
  \begin{center}
    \includegraphics[width=8.5cm]{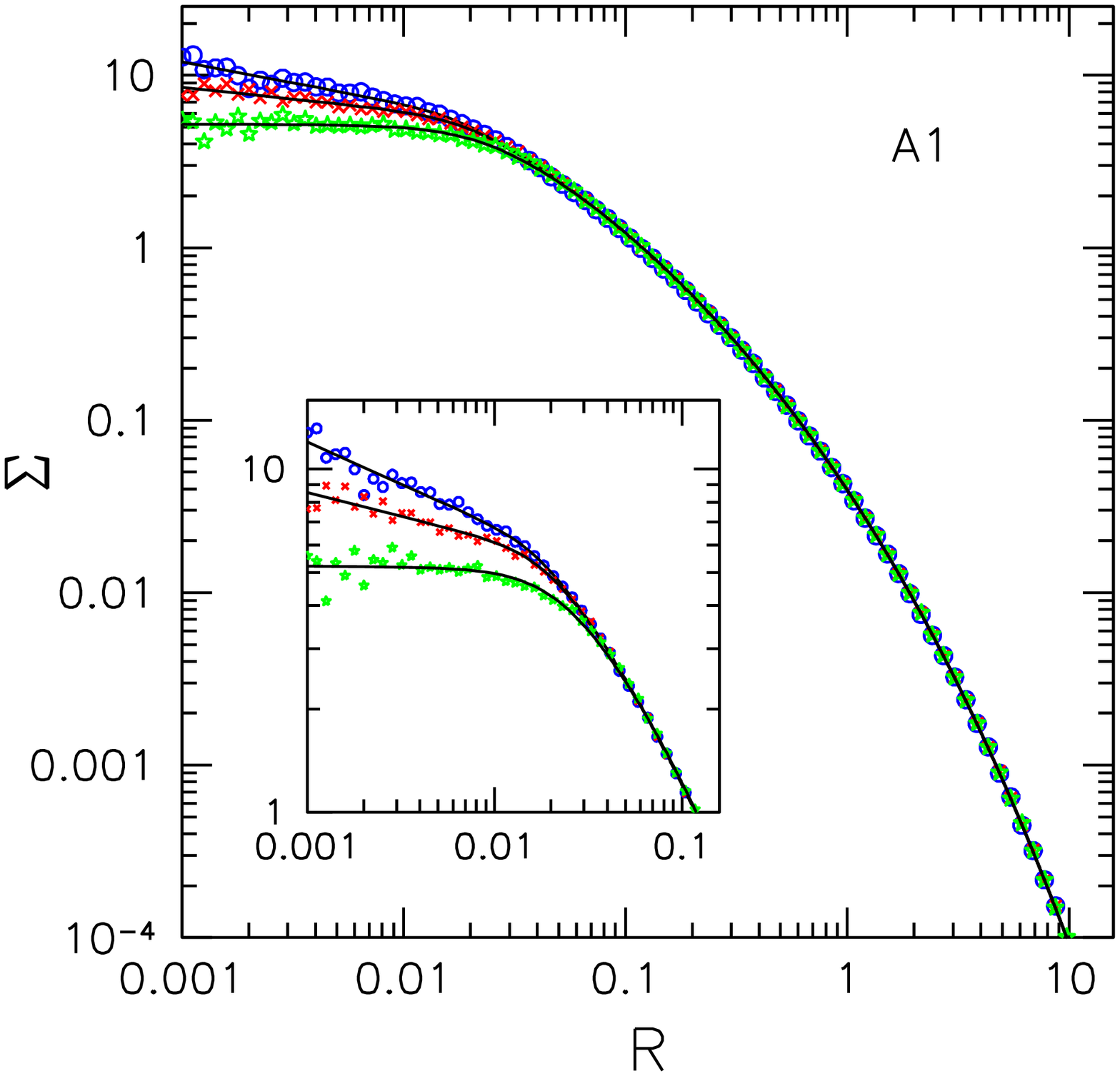}
  \end{center}
  \caption{Projected  density profiles  for model  A1 computed  from the
    $N$-body  data (points), compared with best-fitting core-S{\'e}rsic
    models  (lines), for  three different
    values of the kick velocity ($\vk=0.2,0.4,0.8\,\ve$). 
    The insert shows a  zoom into the central region.}
  \label{fig:fits}
\end{figure}
Three of the best fits for model A1 are shown
in Figure~\ref{fig:fits} (lines) together with the projected density
profiles computed from the $N$-body data (points).

It appears that the host galaxies to recoiling BHs are well
represented by core-S{\'e}rsic profiles.  In particular, the fits
show, once again, that the core tends to expand as the BH oscillates
in and out of it, and that the final core size scales as $r_b \sim
\mbh\,\vk^{\beta}$, with $0.3 \simless \beta \simless 0.6$.  In
addition, the transition from the inner power law to the outer
S{\'e}rsic profile is rather sharp, with best-fit values of $\alpha$
in the range $2\lesssim \alpha \lesssim 7$.  The initial $n=4$ de
Vaucouleurs outer slope is not substantially modified by the BH.

A flattening of the inner profile is also observed in the simulations of
\citet{Boylan:04}, who follow the evolution of a spherical stellar
bulge with a recoiling central black hole using an $N$-body tree code.
They find that the density profile of the system evolves as a consequence
of the gravitational radiation recoil and flattens substantially.
A core of size equal to the BH sphere of influence forms
on a relatively short time-scale, and remains even after several
dynamical times. A flattening of the profile is observed for
recoil velocities smaller and larger than the central escape speed,
though an additional flattening is present if the black hole returns
to the core after the ejection.

A measurable signature of a recoiling BH is the mass deficit, the net
mass removed from the central regions \citep{Milos:02}.  Mass deficits
produced by recoil will add to the depletion caused by the
pre-existing BH binary, which ejects stars from the core during close
encounters.  The deficit produced by the binary is proportional to the
mass of the binary, with only a weak dependence on the mass ratio and
the initial density distribution \citep{Merritt:06}.  Therefore, a
binary BH can only produce a deficit $\mdef \approx \mbh$.  This could
explain the peak in the distribution of observed mass deficits at
$\mdef/\mbh \approx 1$ \citep{Graham:04,ACS:VI}.  The tail of the
distribution, however, extends to values of $\mdef/\mbh \sim 5$.
While such large values might be explained as successive mergers
\citep{Merritt:06}, a recoiling BH represents an interesting
alternative.

\begin{figure}
  \begin{center}
    \includegraphics[width=8.5cm]{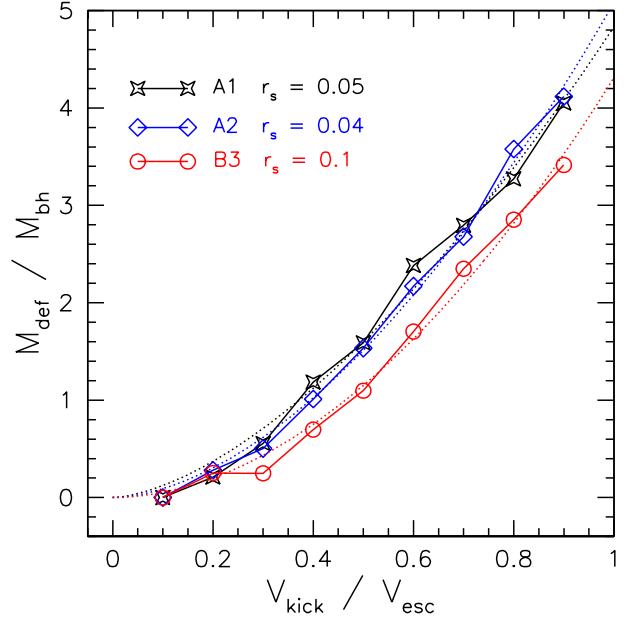}
  \end{center}
  \caption{Mass deficits, as defined in the text, for the different
    runs: A1 (black), A2 (blue), B (red).
    Dashed lines show power-law fits.}
  \label{fig:mdef}
\end{figure}

We evaluated the mass deficits in the final $N$-body models by
computing the difference in stellar mass, enclosed within a sphere of
radius $r_s$, between the initial and final space density profiles.
Given the fact that the deficits depend rather sensitively on the
value of $r_s$, we computed $\mdef$ as a function of $r_s$ for a
number of models and kicks. In all cases, $\mdef$ first increases
rapidly with $r_s$ and then flattens out to an approximately constant
value.  Based on such tests, we concluded that the most appropriate
values to use for the computation of the mass deficits were as
follows: $r_s = 0.05$ for model A1, $0.04$ for model A2 and $0.1$ for
model B.  The results for all three models are shown in
Figure~\ref{fig:mdef}.
Also shown are least-squares fits to $Y=aX^b$, where 
$Y\equiv \mdef/\mbh$ and $X\equiv \vk/\ve$.
The best-fit parameters are:
\begin{eqnarray}
{\rm Model\ A1}:  a &=&4.83,\ b=1.59 \nonumber \\
{\rm Model\ A2}:  a &=&5.08,\ b=1.75 \nonumber \\
{\rm Model\ B\ }: a &=&4.31,\ b=1.90 
\end{eqnarray}

The largest kicks result in mass deficits as large as $4-5\mbh$, which
is consistent with the largest observed deficits \citep{Merritt:06}.
Our definition of mass deficits as the difference in integrated mass
between initial and final profiles implies that our estimates do not
take into account any depletion prior to the kick.  One should
therefore add the contribution from the binary evolution phase ($\mdef
\approx 1 \mbh$) to our measured deficits before comparing with the
observed values.

The sensitivity of $\mdef$ to $r_s$, which presumably is a feature of
real luminosity profiles as well, suggests that a more objective way
be found to measure mass deficits.

We compare the projected density profiles obtained from the $N$-body
simulations to the brightness profiles of a sample of early-type
galaxies in the Virgo cluster observed with the {\it Advanced Camera
for Surveys} (ACS) on the Hubble Space Telescope \citep{ACS:VI}.  In
this study, the authors find that, while simple S{\'e}rsic models
generally provide a good representation of the global galaxy profiles,
the brightest galaxies require a power-law component within a
characteristic break radius and are therefore best modeled with
core-S{\'e}rsic profiles.  

We select two representative galaxies in the sample and compare their
surface density profiles with each of the 27 final profiles obtained
from the simulations (9 values of the kick velocities $0.1\dots0.9$
for each of the 3 models A1, A2, B).

The brightest Virgo galaxy, VCC1226 (M49, NGC 4472), has the largest
value of mass deficit ($\mdef/\mbh \sim 4$) and has a S{\'e}rsic index
that is not too far from our $N$-body models, $n \sim
5.9$\footnote{Most of the bright galaxies in the ACS sample have $n
\simgreat 7$.}. On the other hand, VCC 731 (NGC 4365) has a relatively
small core and a typical mass deficit of $\sim 1\mbh$
\citep{Merritt:06}.  For each galaxy, we scale the $N$-body profiles
to have the same $r_b$ and $\Sigma (r_b)$ as the galaxy itself.

\begin{figure}
  \begin{center}
    \includegraphics[width=8.5cm]{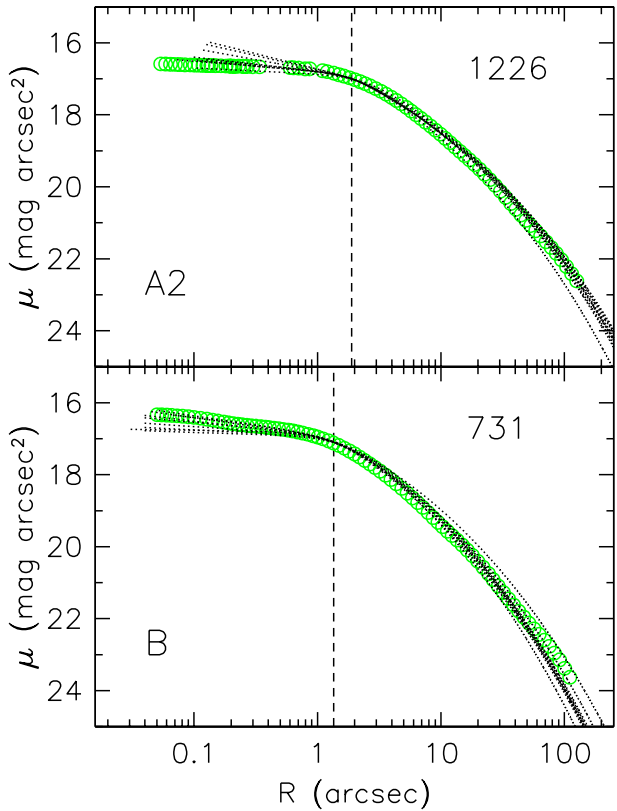}
  \end{center}
  \caption{Surface brightness profiles for the Virgo galaxies VCC 1226
    (top) and VCC 731 (bottom) from the ACS sample compared to the
    $N$-body profiles obtained from the best fitting of the three
    models. The different lines correspond to the 9 different kicks
    $\vk/\ve = 0.1\dots0.9$.}
  \label{fig:sbp}
\end{figure}
Figure~\ref{fig:sbp} shows that the brightness profiles of both
galaxies can be reasonably well fit by (at least) one of the $N$-body
models.  In particular, the profile of VCC 1226 is well fit by models
with $\vk \ge 0.4 \ve \approx 550\kms$ while VCC 731 is well fit by
models with $\vk \simgreat 0.1 \ve \approx 110\kms$.
This indicates that observed brightness profiles, and
even the largest cores, can be well reproduced by the gravitational
recoil kicks.

\section{Evolution times in real galaxies}
\label{sec:times}

Given a galaxy's effective radius $\re$ and total mass $M_{\rm gal}$,
equations~(\ref{eq:units}) relate our $N$-body units to physical
units.  We adopt the scaling relations derived from the ACS Virgo
cluster survey of \cite{ACS:I} between $\re$ and absolute blue
magnitude $M_B$ for early-type galaxies.  \cite{ACS:VI} found, for
Virgo E galaxies fainter than $M_B\approx -20.5$, a mean relation 
\begin{equation}
\log_{10}\re = 0.144 - 0.05 \left(M_B+20\right) 
\end{equation} 
where $\re$ is in kpc. (Brighter galaxies obey a different relation and are considered
separately below.)  We relate $M_B$ to galaxy mass using Gerhard et
al.'s (2001) expression for the mass to light ratio in the blue band:
\begin{equation} 
\log_{10}\left[\left(\frac{M}{L}\right)/ \left(\frac{M}{L}\right)_\odot\right]_B \approx 1.17 + 0.67\log_{10}\left(\frac{L_B}{10^{11}\,L_{\odot,B}}\right).
\label{eq:moverl}
\end{equation}
Equation~(\ref{eq:moverl}) was derived from dynamical
modeling of galaxies with $M_B\gtrsim -22.5$
and represents an average for the matter within the effective radius,
including dark matter if present.
Combining these relations gives
\begin{equation}
\re \approx 1.2\ {\rm kpc} \left(\frac{\mgal}{10^{10}\msun}\right)^{0.075}.
\label{eq:revsmb}
\end{equation} The dependence of $\re$ on $\mgal$ is weak, a consequence of the
low slope of the $\re-M_B$ relation.  However we note that the scatter
in this relation is large (e.g. \citealp{ACS:VI}, Fig.~136).
 
Some fiducial values, and their implied $N$-body scalings
(from equation~\ref{eq:units}), are:
\begin{eqnarray}
\mgal=3\times 10^9\msun\ \ && \ \ \re=1.1\ {\rm kpc} \nonumber \\
&&\left[T\right] = 1.0\times 10^7 {\rm yr} \ \ \ \  
\left[V\right] = 110\kms, 
\nonumber \\
\mgal=3\times 10^{10}\msun\ \ && \ \ \re=1.3\ {\rm kpc} \nonumber \\
&&\left[T\right] = 4.1\times 10^6 {\rm yr} \ \ \ \ 
\left[V\right] = 315\kms, 
\nonumber \\
\mgal=3\times 10^{11}\msun\ \ && \ \ \re=1.5\ {\rm kpc} \nonumber \\
&&\left[T\right] = 1.7\times 10^6 {\rm yr} \ \ \ \ 
\left[V\right] = 910\kms.
\nonumber
\label{eq:scalings}
\end{eqnarray}
The trend of decreasing $\left[T\right]$ with increasing $\mgal$
reflects the well-known higher density of more massive galaxies
\citep{Graham:03}.  The central escape velocities in our models are
$2.0\lesssim\ve\lesssim 2.2$ in $N$-body units, corresponding to
$\ve\approx 2.1\times\left[V\right] \approx 2000\kms$ when scaled to a
$3\times 10^{11}\msun$ galaxy.  This agrees well with escape
velocities of bright E-galaxies derived from more detailed modeling
(e.g. Fig.~2, \citealp{MMFHH:04}).

\begin{figure}
  \begin{center}
    \includegraphics[width=7.cm]{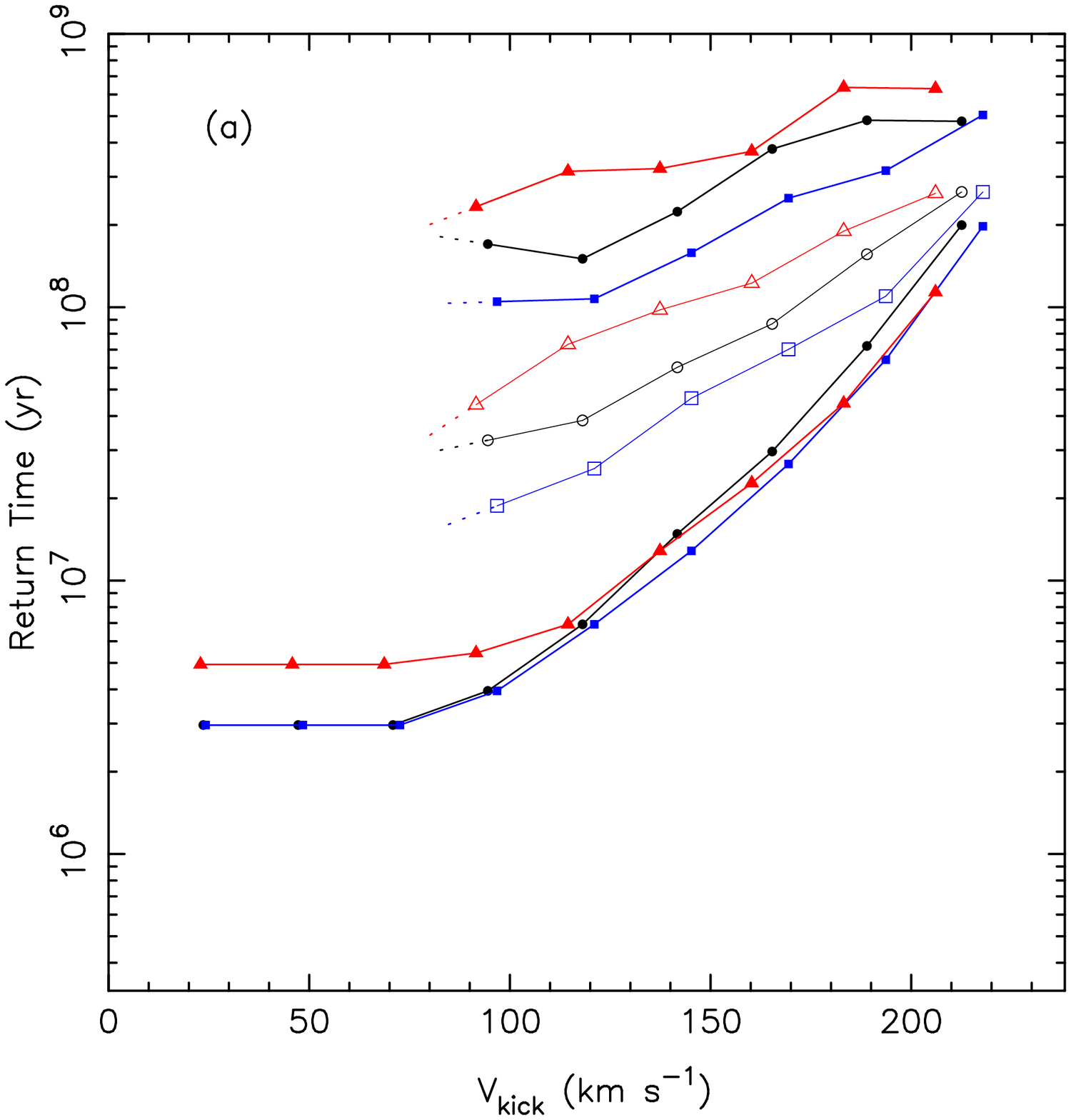}
    \includegraphics[width=7.cm]{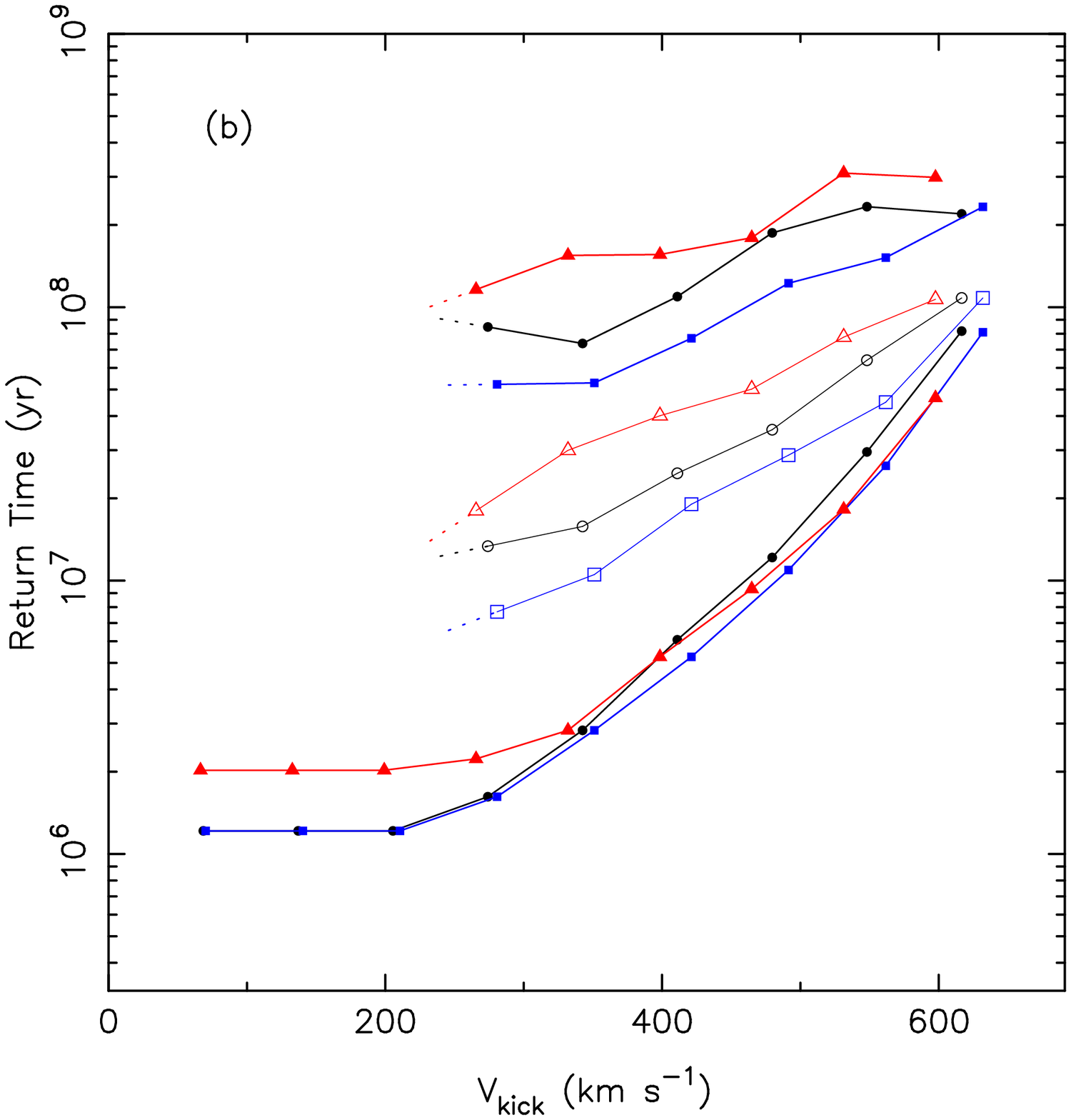}
    \includegraphics[width=7.cm]{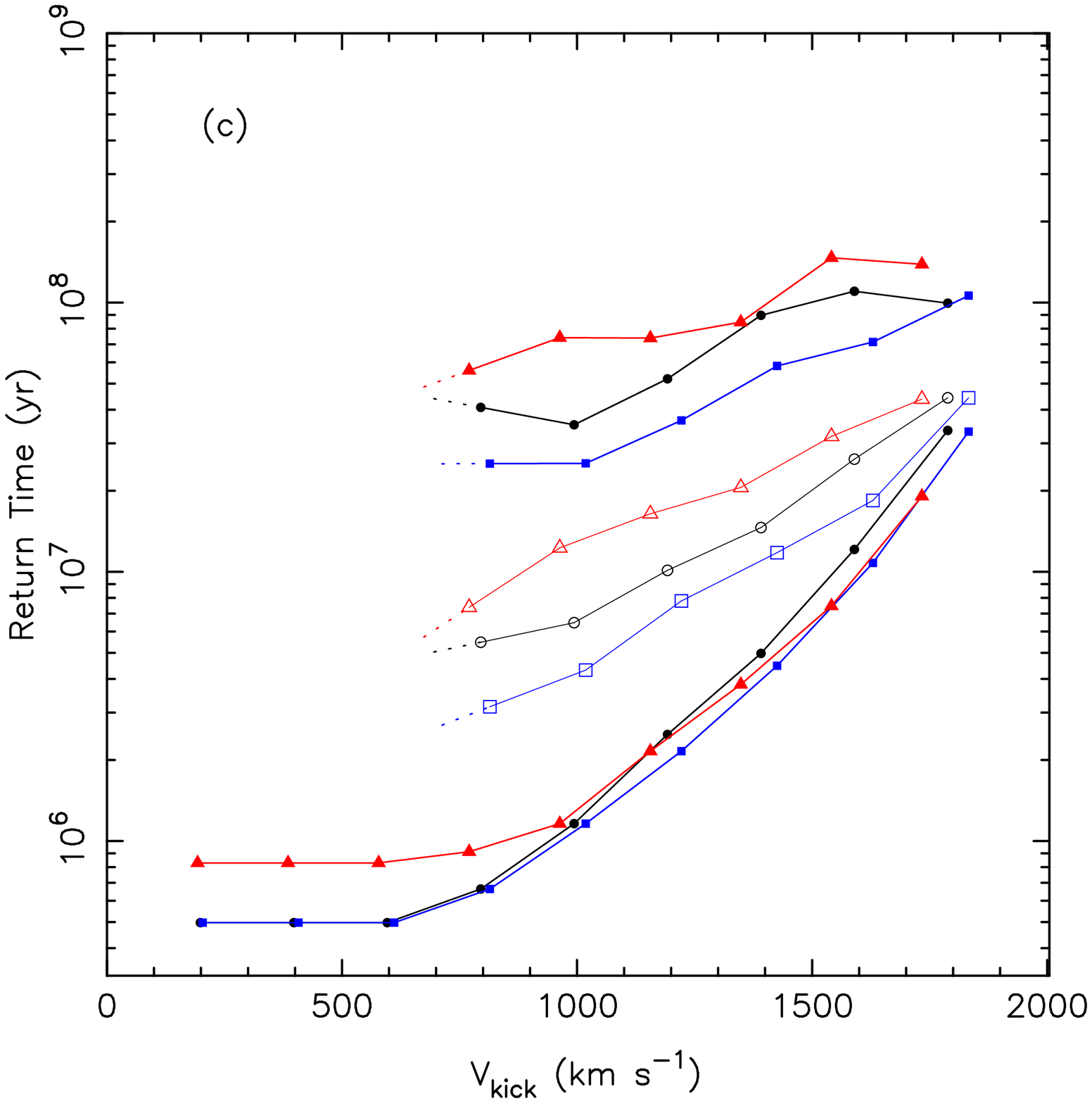}
  \end{center}
  \label{fig:times1}
  \caption{Return times of kicked BHs.
    These are plots of the $N$-body values given in
    Table~\ref{tab:times}, scaled to physical units
    using equations~(\ref{eq:units}) and~(\ref{eq:revsmb}).
    {\it Lower (filled) symbols:} $T_I$;
    {\it Middle (open) symbols:} $T_I+T_{II}$, with $T_{II}$ taken
    directly from the simulations;
    {\it Upper (filled) symbols:} $T_I+T_{II}$, with $T_{II}$ 
    scaled to $N_{\rm gal}$ using equation~(\ref{eq:addtime}).
    (a) $\mgal=3\times 10^9\msun$; 
    (b) $\mgal=3\times 10^{10}\msun$; 
    (c) $\mgal=3\times 10^{11}\msun$.
  }
  \label{fig:times}
\end{figure}

All of the times listed in Table~\ref{tab:times} can be scaled to
physical units using these relations.  Figure~\ref{fig:times} shows
the result for the three fiducial values of $\mgal$.  We have used
equation~(\ref{eq:addtime}) to correct the measured $T_{II}$ values to
different values of $N_{\rm gal}\equiv \mgal/m_\star$ assuming
$m_\star=\msun$; we also show, as conservative lower limits, the
$T_{II}$ times obtained directly from the simulations.

Figure~\ref{fig:times} seem to suggest that return times depend
discontinuously on $\vk$, since Phase II does not appear to exist in
our simulations when $\vk\lesssim 0.3\ve$.  As discussed above, this might
not be true in simulations with larger $N$, or in real galaxies.  In
any case, for $\vk\gtrsim 0.4\ve$, return times are dominated by the
time spent in Phase II (``BH-core oscillations'').

The brightest galaxies, $M_B\lesssim -21$, appear to obey a different
scaling relation between $\re$ and $M_B$ than the
relation~(\ref{eq:revsmb}) given above \citep{ACS:VI}.  Furthermore,
these bright galaxies are typically fit by S{\'e}rsic indices in the
range $5\lesssim n \lesssim 10$, larger than the value $n=4$ adopted
here for the $N$-body models.  On the other hand, the brightest E
galaxies often have resolved cores with well-determined sizes and
densities (cf. \S\,\ref{sec:profiles}).  Furthermore,
equation~(\ref{eq:tauomega}) gives the damping time $\tau$ in Phase II
in terms of core properties alone.

We define the Phase II return times for these galaxies
as the time for the BH's energy to decrease from
\begin{equation}
\frac{1}{2} \omega_c^2 r_c^2 \approx \frac{2}{3} \pi G \rho_c r_c^2,
\end{equation}
the BH's energy when it first re-enters the core, to
\begin{equation}
\frac{1}{2} V_{\rm brown}^2 \approx \frac{3}{2}\frac{m_\star}{\mbh}\sigma_c^2,
\end{equation}
the Brownian energy,
assuming an energy damping time constant of $\tau$;
for the latter we take equation~(\ref{eq:tauomega}).
This time is 
\begin{mathletters}
\begin{eqnarray}
T_{II} &=& \cal{N} \tau, \nonumber \\
\tau &\approx & 15 \frac{\sigma_c^3}{G^2 \rho_c\,\mbh}
\label{eq:vt1} \\
&\approx &1.2\times 10^7{\rm yr} 
\left(\frac{\sigma_c}{250\kms}\right)^3
\left(\frac{\rho_c}{10^3\msun\,{\rm pc}^{-3}}\right)^{-1}
\left(\frac{\mbh}{10^9\msun}\right)^{-1}, \nonumber \\
\cal{N} &=& \ln\left(\frac{1}{F^2}\frac{\mbh}{m_\star}\right) \nonumber \\
&\approx& \ln\left(\frac{1}{F^2} \frac{\mbh}{\mgal} \frac{\mgal}{\msun}\right),
\label{eq:vt2}
\end{eqnarray}
\end{mathletters}
with $F\approx 2$ the form factor defined above,
and we have again assumed $m_\star=\msun$.

\begin{figure}
  \begin{center}
    \includegraphics[width=8.5cm]{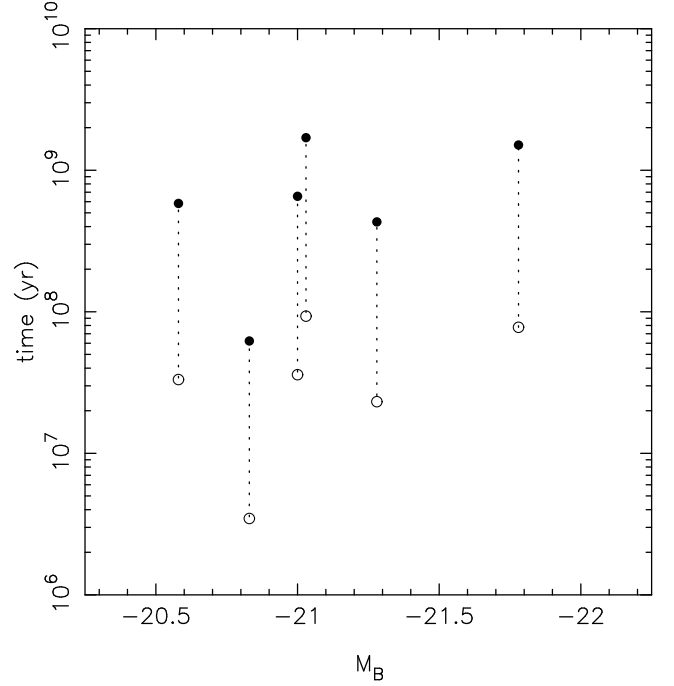}
  \end{center}
  \caption{Estimates of the return time in Phase II for supermassive
    black  holes in the six brightest Virgo galaxies, excluding
    M87 \citep{ACS:I}.
    This plot assumes that the SMBHs have received a kick
    large enough to remove them from the core initially.
    Lower (open) symbols show the energy decay time constant
    in the core, $\tau$ (equation~\ref{eq:vt1}), 
    while upper (filled) symbols show $\cal{N}\tau$, 
    where $\cal{N}$ is the estimated number of time constants
    required for the BH's velocity
    to decay to the Brownian value (equation~\ref{eq:vt2}).
  }
  \label{fig:virgotimes}
\end{figure}

Figure~\ref{fig:virgotimes} shows estimates of $\tau$ and
$\cal{N}\tau$ for the six brightest galaxies in the ACS Virgo sample
excluding M87, which has an active nucleus \citep{ACS:I}.  Of course,
this figure is only meaningful under the assumption that the BHs in
these galaxies have received large enough kicks to remove them
completely from the core, i.e. $\vk\approx 10^3\kms$.  But if this did
occur, Figure~\ref{fig:virgotimes} suggests that return times would be
of order $1$\,Gyr.  Such a long time is comparable with the mean time
between galaxy mergers in a dense environment like the Virgo cluster.
Hence, a SMBH might never return fully to the center before another
SMBH spirals in.

\section{Observable Consequences}
\label{sec:obs}

\subsection{Likelihood of Large Kicks}
\label{sec:kicks}
Kicks large enough to remove SMBHs from cores,
$\vk\gtrsim 0.4\ve$,  range from 
$\sim 90\kms$ for $\mgal=3\times 10^9\msun$, 
to $\sim 750\kms$ for $\mgal=3\times 10^{11}\msun$, 
to $\sim 1000\kms$ for $\mgal=3\times 10^{12}\msun$, 
based on the fiducial scalings in \S\,\ref{sec:times}.
The most propitious configuration for the kicks appears to be
an equal-mass binary in which the individual
spin vectors are oppositely aligned and oriented parallel to the
orbital plane \citep{Campanelli:07b,Gonzalez:07b}.
Assuming this most favorable orientation, 
and setting the spins to their maximal values, 
the maximum kick (oriented parallel to the binary angular
momentum vector) is believed to scale with binary mass ratio 
$q\equiv M_2/M_1\le 1$ as
\begin{equation}
V_{\rm max} \approx 6\times 10^4 {\rm km\ s}^{-1} {q^2\over (1+q)^4}
\end{equation}
\citep{Lousto:07}.
Mass ratios as small as $q\approx 0.2$ can therefore
result in kicks $\gtrsim 1000$ km s$^{-1}$.
While the assumption of near-maximal spins is probably not an extreme
one (e.g. \citealp{Shapiro:05,Gammie:04}),
orienting the BHs with their spins perpendicular to the orbital
angular momentum may seem odd, particularly in gas-rich galaxies
\citep{BRM:07}.
However there is considerable circumstantial evidence that SMBH 
spin axes bear no relation
to the orientations of the gas disks that surround them
\citep{Kinney:00,Gallimore:06,Borguet:07}
and this is presumably even more true with
respect to the directions of infalling BHs in gas-free galaxies.
If SMBH spins do orient parallel with orbital angular momenta,
the maximum kick is more modest and contains contributions
from both the ``mass asymmetry'' ($M_1\ne M_2$) and from the 
spins.
The two kick components, both of which are parallel to the orbital plane,
are believed to be approximately independent
and to scale roughly as
\begin{mathletters}
\begin{eqnarray}
V_{\rm mass} &\approx& V_1{q^2(1-q)\over (1+q)^5}, \\
V_{\rm spin} &\approx& V_2{q^2\over (1+q)^5}\left(\alpha_2 - q\alpha_1\right);
\label{eq:V0}
\end{eqnarray}
\end{mathletters}
$V_1\approx V_2\approx 10^4$ km s$^{-1}$ and
$\alpha$ denotes a dimensionless spin, $-1\le\alpha\le 1$
\citep{Lousto:07,Baker:07,LZ:07}.
$V_{\rm mass}$ peaks at $\sim 200$ km s$^{-1}$
for $q\approx 0.4$ while $V_{\rm spin}$ peaks at $\sim 600$ km s$^{-1}$
for $q\approx 1, \alpha_1=-\alpha_2=1$.
In this less-favorable configuration, kicks could remove 
SMBHs only from the cores of low-to-moderate luminosity galaxies.
Estimates of the kick velocity distribution (e.g. \citealp{SB:07})
are extremely uncertain since they depend on the unknown distributions
of SMBH mass ratios, spins and spin orientations.
In what follows, we will focus on the consequences of kicks
that are large enough to remove SMBHs from galaxy cores
and to excite the long-lived oscillations that we described
above.

\subsection{Offset and Double Nuclei}
\label{sec:nuclei}

\cite{Lauer:05} identified five galaxies in which the point
of maximum surface brightness is displaced from the
center of the isophotes defined by the galaxy on large scales.
All are luminous, ``core'' galaxies.
Contour plots for two of the galaxies, NGC 507 and 1374 
(Figs.~17, 18 of \citealt{Lauer:05}), look strikingly similar
to the ``Phase II'' isodensity plots in Figure~\ref{fig:cont}.
Displacements are cited for NGC 507 ($0''.06\approx 19\rm pc$),
NGC 1374 ($0''.02\approx 2.1\rm pc$), and
NGC 7619 ($0''.04\approx 11\rm pc$), all of which are of
order the core radii in these galaxies.
The five galaxies with offset nuclei comprise 12\% of the 
Lauer et al. ``core'' galaxy sample;
no offset nuclei were found among the ``power-law'' 
(non-cored) galaxies.
Several of the offsets are close to the resolution
limit, and some offsets might go unobserved due to
projection, so it is likely that offset nuclei are
quite common in ``core'' galaxies.
If the offsets are produced by oscillations like those
in Figure~\ref{fig:cont}, the SMBHs in these galaxies would be
located on the opposite side of the galaxy photocenter
from the point of peak brightness.
Phase II oscillations can also produce a ``double nucleus''
morphology (e.g. Figure~\ref{fig:cont}, frame 8) with the BH
located at either the higher or secondary peak.
This is a reasonable model for the double nucleus in NGC 4486B
\citep{Lauer:96}, since the two peaks are closely matched
in brightness and are offset by similar amounts ($\sim 6\rm pc$)
from the galaxy photocenter.
Galaxies with central minima in the surface brightness
(e.g. NGC 4406, NGC 6876; \cite{Lauer:02}) might also be
explained in this way.
This model is probably not as appropriate for the more famous
double nucleus in M31, since M31 is not a ``core'' galaxy,
and one of the brightness peaks (the one associated
with the SMBH) lies close to the galaxy photocenter \citep{Lauer:93}.

\subsection{Displaced AGN}
\label{sec:AGN}

An ejected SMBH can appear as a spatially or kinematically
displaced AGN \citep{Kapoor:76,Kapoor:83a,Kapoor:83b}.
A recoiling SMBH retains gas that is orbiting around it
within a distance
\begin{equation}
r_{\rm eff} \approx \frac{G \mbh}{\vk^2} \approx
0.5\,{\rm pc}\ M_8 V_{\rm k, 1000}^{-2}
\end{equation}
with $M_8\equiv\mbh/10^8\msun$ and $V_{\rm k, 1000}\equiv \vk/1000\kms$.
An accretion disk if present would mostly be retained,
and for kicks $\lesssim 10^3\kms$, $r_{\rm eff}$ is large enough
to encompass most of the broad emission-line region gas as well.
Narrow emission lines originate in gas moving in the gravitational
potential of the host galaxy and would not follow a recoiling SMBH
\citep{Merritt:HE0450}.
\cite{Bonning:07} used this argument to search for kinematic
offsets between spectral features associated with the
broad- and narrow emission line regions.
No convincing cases were found.
This may be a consequence of the rapid decrease in SMBH energy 
during Phase I (Fig.~\ref{fig:eoft}).
In the longer-lived oscillations that characterize Phase II,
the rms velocity of the SMBH drops from
\begin{equation}
\sim 90\kms 
\left(\frac{\rho_c}{10^3\msun\ {\rm pc}^{-3}}\right)^{1/2}
\left(\frac{r_c}{30\,{\rm pc}}\right)
\end{equation}
when it first re-enters the core, to
\begin{equation}
\sim 0.03\kms \left(\frac{\mbh}{10^8\msun}\right)^{-1/2}
\left(\frac{\sigma_c}{200\kms}\right)
\end{equation} 
in the Brownian regime.
Such small velocity offsets would be difficult to detect.
An alternative approach would be to search for linear
displacements $\Delta R$ between the AGN emission and the 
peak of the stellar surface brightness.
This displacement is $\sim r_c$ at the start of Phase II,
dropping to $\sim \sqrt{m_\star/\mbh}r_c$ in the Brownian
regime; the exponential nature of the damping implies 
an approximately uniform distribution of 
$\ln\Delta R$ during Phase II.
Relatively large ($\sim 10-100 \rm pc$) offsets between the AGN and either the
stellar density peak or the center of rotation have in fact
been claimed in a number of galaxies based on integral-field
spectroscopy \citep[e.g.][]{Media:93,Media:05}.

\subsection{Wiggling Jets}
\label{sec:jets}

During Phase II, the SMBH oscillates sinusoidally within the
core with roughly constant period,
\begin{equation}
\frac{2\pi}{\omega_c} \approx 1.4\times 10^6 \,{\rm yr} 
\left(\frac{\rho_c}{10^3\msun\ {\rm pc}^{-3}}\right)^{-1/2},
\end{equation}
and with velocities as given above.
Such motion will induce periodic deviations in the 
velocity and direction of a jet emitted by the SMBH
\citep{Roos:92}.
If the jet is oriented perpendicularly to the direction of
motion of the SMBH, the jet direction is fixed, and the
jet material moves on a cylindrical surface with radius
equal to the radius of the SMBH's orbit.
If the jet velocity has some component parallel to the 
SMBH's motion, the two velocities add and the cylinder becomes 
a cone over which the jet precesses \citep{Roos:93}.
Such models have been used to explain the helical distortions
observed in a number radio sources;
the inferred orbital periods are typically $1-100\,\rm yr$,
and the jet accelerations are usually ascribed to the orbit of the
jet-producing SMBH around a second SMBH in a close 
($\ll 1 \rm pc$) binary pair.
However some sources are fit by models with longer periods.
For instance, the morphology of the {\bf C}-type source 
3C 449 has been reproduced assuming jet forcing
with a period of $\sim 10^7\,\rm yr$ \citep{Hardee:94}.
Such long periods are sometimes explained in terms of
bulk motion of the galaxy hosting the radio source
\citep{Icke:78},
but oscillations of the SMBH within the core might
provide a tenable alternative in some cases.

\subsection{Oversized Cores and Hypermassive Black Holes}
\label{sec:hyper}

Cores generated by kicked SMBHs can be substantially larger 
than those produced by ``core scouring'' from a binary SMBH 
\citep{MM:01,Merritt:06}, particularly when 
$\vk\gtrsim 0.4\ve$.
As shown in \S\,\ref{sec:profiles} (Figures~\ref{fig:density}-\ref{fig:sbp}),
kick-induced cores can be as large as those
observed in some of the brightest ``core'' galaxies,
having mass deficits of $4-5\mbh$ and core radii 
several times the SMBH influence radius, 
or $\sim 5\%$ of the galaxy's half-light radius.
(Similar conclusions were reached already by 
\cite{Boylan:04} and \cite{MMFHH:04}.)
While the majority of observed mass deficits lie in the
range $0.5\lesssim \mdef/\mbh\lesssim 1.5$, 
some E galaxies have $\mdef/\mbh\gtrsim 3$,
too large to be easily explained by core scouring.
\cite{Lauer:07} invoked the oversized cores, along with other
circumstantial evidence, to argue that the SMBHs in the brightest
E galaxies are ``hypermassive,'' $\mbh\gtrsim 10^{10}\msun$.
An alternative possibility is that the largest cores have been
enlarged by kicks.
Figure~\ref{fig:mdef}, combined with earlier $N$-body results
\citep{Merritt:06}, suggests that the total mass deficit
generated by a binary SMBH following a single galaxy merger is
\begin{eqnarray}
\mdef &=&M_{\rm def,bin} + M_{\rm def,kick} \nonumber \\
M_{\rm def,bin} &\approx& 0.7 q^{0.2}\mbh, \nonumber \\
M_{\rm def,kick} &\approx& 5\mbh \left(\vk/\ve\right)^{1.75}
\label{eq:mdefofv}
\end{eqnarray}
where $M_{\rm def,bin}$ and $M_{\rm def,kick}$ are the mass deficits 
generated by ``core scouring'' and by the kick respectively  and
$q\equiv M_2/M_1\le 1$ is the binary mass ratio.
It has been argued \citep{Merritt:06} that the ratio
$M_{\rm def,bin}/\mbh$ increases in multiple mergers, 
and the same is likely to be
true for kick-induced core growth.
Thus, the decrease in typical values of of $\vk/\ve$ with 
increasing galaxy luminosity might be offset by the greater 
number of mergers that contribute to the growth of luminous
galaxies, leading to comparable values of $\mdef/\mbh$.
In any case, the possibility that core growth is dominated
by the kicks should be considered in future studies.

\acknowledgements
We thank D. Axon, M. Campanelli, S. Komossa, C. Lousto, R. Miller,
A. Robinson and Y. Zlochower for illuminating discussions.  This work
was supported by grants AST-0206031, AST-0420920 and AST-0437519 from
the NSF, grant NNG04GJ48G from NASA, and grant HST-AR-09519.01-A from
STScI.

\bibliographystyle{apj}
\bibliography{biblio}

\end{document}